\documentclass[onecolumn,aps,pra,reprint,superscriptaddress,longbibliography]{revtex4-1}

\usepackage{amsmath}
\usepackage{graphicx}
\usepackage[lofdepth,lotdepth, caption=false]{subfig}
\usepackage{verbatim}
\usepackage{color}
\usepackage{dcolumn}
\usepackage{bm}
\usepackage{miller}
\usepackage{color}
\usepackage{amsfonts}
\usepackage{setspace}
\usepackage{xr-hyper}
\usepackage{hyperref}
\usepackage[english]{babel}

\newcommand{\beginsupplement}{%
        \setcounter{table}{0}
        \renewcommand{\thetable}{S\arabic{table}}%
        \setcounter{figure}{0}
        \renewcommand{\thefigure}{S\arabic{figure}}%
     }

\begin{document}

\title{Magnetic dynamics in NiTiO$_3$ honeycomb antiferromagnet using neutron scattering}

\author{Srimal Rathnayaka}
\affiliation{Department of Physics, University of Virginia, Charlottesville,
Virginia 22904, USA}

\author{Luke Daemen}
\affiliation{Neutron Scattering Division, Oak Ridge National Laboratory, Oak Ridge, Tennessee 37831, USA}

\author{Tao Hong}
\affiliation{Neutron Scattering Division, Oak Ridge National Laboratory, Oak Ridge, Tennessee 37831, USA}

\author{Songxue Chi}
\affiliation{Neutron Scattering Division, Oak Ridge National Laboratory, Oak Ridge, Tennessee 37831, USA}

\author{Stuart Calder}
\affiliation{Neutron Scattering Division, Oak Ridge National Laboratory, Oak Ridge, Tennessee 37831, USA}

\author{John A.~Schneeloch}
\affiliation{Department of Physics, University of Virginia, Charlottesville,
Virginia 22904, USA}

\author{Yongqiang Cheng}
\affiliation{Neutron Scattering Division, Oak Ridge National Laboratory, Oak Ridge, Tennessee 37831, USA}

\author{Bing Li}
\affiliation{Neutron Scattering Division, Oak Ridge National Laboratory, Oak Ridge, Tennessee 37831, USA}

\author{Despina Louca}
\thanks{Corresponding author}
\email{louca@virginia.edu}
\affiliation{Department of Physics, University of Virginia, Charlottesville,
Virginia 22904, USA}

\begin{abstract}

The ilmenite NiTiO$_3$ consists of a buckled honeycomb lattice, with the Ni spins aligned ferromagnetically in-plane and antiferromagnetically out-of-plane. Using neutron spectroscopy, the magnetic structure and the dynamics were investigated as a function of temperature. Dispersive acoustic bands and nearly dispersionless optical bands at $\approx$3.7 meV are described by a highly anisotropy Heisenberg model with stronger antiferromagnetic (AFM) out-of-plane, weaker ferromagnetic (FM) in-plane interactions and an anisotropy gap of 0.95 meV. The order parameter yields a critical exponent between the Heisenberg and two-dimensional Ising models, consistent with highly anisotropic Heisenberg systems. The frustration parameter $\approx$ 2 supports a weakly  frustrated system.

\end{abstract}

\maketitle

\section{Introduction}
\label{introduction}

Interest in low-dimensional magnetic materials (LDMs) has been growing in part due to their widespread application in magnetic memory devices and magnetic sensors\cite{soumyanarayanan2016emergent,miao20182d,fert2008nobel,wolf2001spintronics,felser2007spintronics}, and also because they are prone to phenomena such as topological and, frustrated magnetism, and multiferroic behaviors\cite{tokura2019magnetic,vsmejkal2018topological,bishop2012frustrated,hu2019progress,dey2021magnetostructural}. Specifically, the halides CrX$_3$ (X = Cl, Br, I), sulfides APS$_3$ (A = Mn, Fe, Co, Ni) and ilmenites ATiO$_3$ consist of layered honeycomb magnetic structures where their reduced dimensionality can lead to Dzyaloshinskii–Moriya interactions (DMI), quantum hall effect, spin Nernst effect, and Dirac magnons to mention a few ~\cite{schneeloch2022gapless,chen2023damped,ederer2008electric,niu2017quantum,haldane1988model,go2024magnon,mcclarty2022topological,yuan2020dirac}. For instance, gapped and gapless Dirac magnons were elucidated in CrI$_3$, CrCl$_3$ and CoTiO$_3$ using inelastic neutron scattering~\cite{schneeloch2024antiferromagnetic, schneeloch2022gapless,rathnayaka2024temperature}. 

In particular, the ilmenite CoTiO$_3$ exhibits strong spin-orbit coupling that yields an effective spin S = $\frac{1}{2}$, XXZ-type magnetic interactions and Dirac-type magnon dispersions. On the other hand, isostructural systems such as MnTiO$_3$ and FeTiO$_3$ exhibit three-dimensional (3D) magnetic behaviors with relatively strong interplane coupling, unlike in CoTiO$_3$, that exhibits magnetic behavior with weaker interplane interactions \cite{goodenough1967theory,osmond1964magnetic,shirane1959neutron,kato1983coexistence,hwang2021spin,yuan2020dirac,yuan2020spin,yuan2024field,elliot2021order,dey2021magnetostructural,rathnayaka2024temperature,newnham1964crystal,yamaguchi1986re,charilaou2012large}. Included in the ilmenite family is NiTiO$_3$ that shares a similar magnetic structure as in CoTiO$_3$. However, NiTiO$_3$ exhibits very different magnetic dynamics, absent of Dirac magnon nodes and topological properties. Earlier studies on the dielectric properties showed a frequency response that is significant at high temperatures~\cite{acharya2015structural,bamzai2011dielectric,zhang2015structural} but little is known of the magnetic properties of NiTiO$_3$ and is the focus of the present work.

\begin{figure}[b]
\begin{center}
\includegraphics[width=8.6cm]{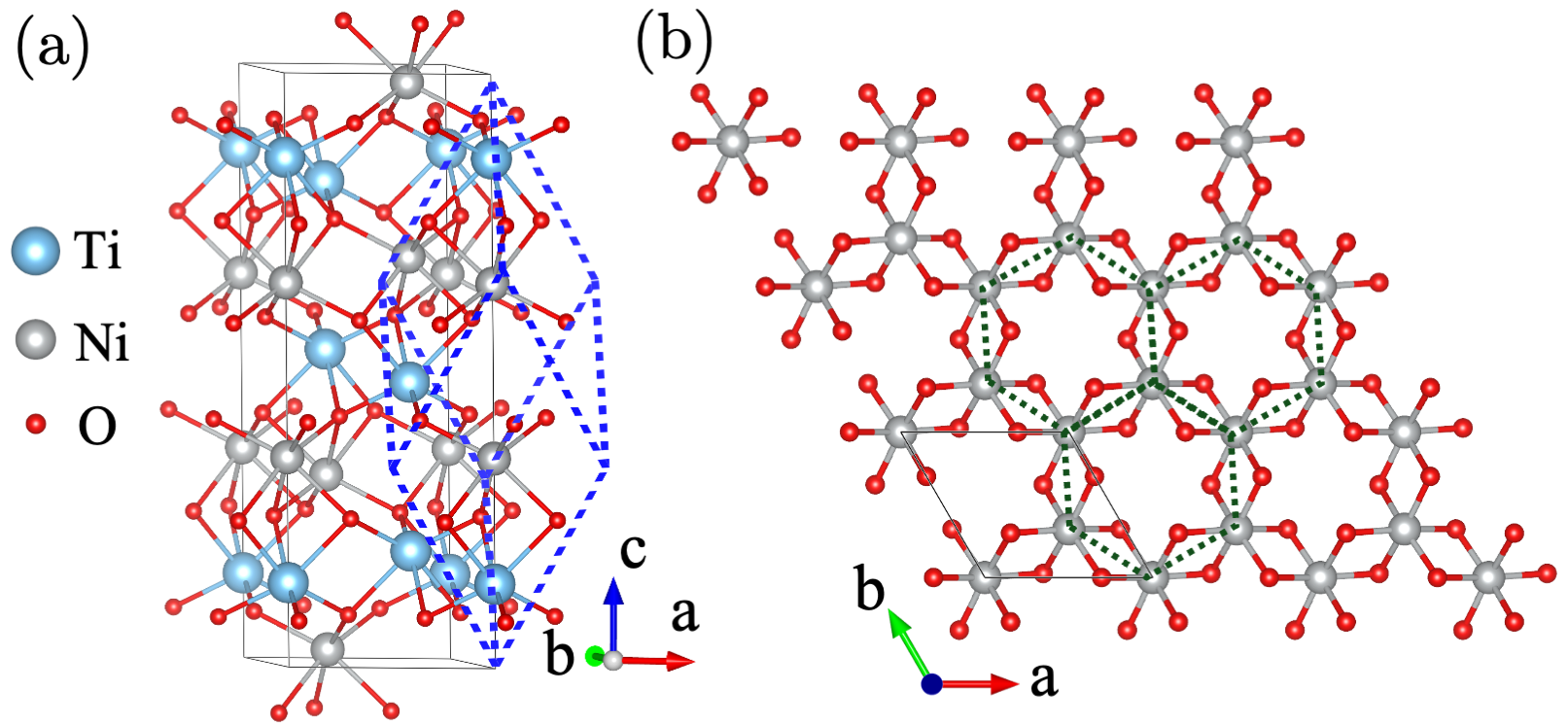}	
\end{center}
\caption{(a) The unit cell of NiTiO$_3$ with the R$\overline{3}$ symmetry. The blue dotted rhombohedron is the unit cell in rhombohedral representation.
(b) In-plane honeycomb-like structure of the Ni atoms. The dashed line represent the honeycomb-like unit in the system.} 
\label{fig1}
\end{figure}

The NiTiO$_3$ crystal structure is rhombohedral with the R$\overline{3}$ symmetry and the unit cell shown in Fig. 1(a) \cite{goodenough1967theory,osmond1964magnetic}. The unit cell consists of alternating layers of corner-sharing NiO$_6$ and TiO$_6$ polyhedra in a trigonal prismatic coordination. The layers form a honeycomb-like structure within the $ab$-basal plane, as shown in Fig. 1(b). NiTiO$_3$ orders antiferromagnetically with a Neel transition temperature (T$_N$), 23 K, with spin $S$=1, and an effective orbital angular momentum in the ground state that is quenched, unlike in some other ilmenites \cite{goodenough1967theory}. The spin structure of the system is A-type, as illustrated in the inset of Fig. 2(b). The spins are aligned along the $ab$-plane, with in-plane FM interactions and  out-of-plane AFM interactions, similar to the spin structure observed in CoTiO$_3$ ~\cite{goodenough1967theory}. The propagation wave vector, $\textit{k}$=(0, 0, 3/2) and the reported magnetic moment per Ni atom is 1.46(1) $\mu_B$ \cite{dey2021magnetostructural,dey2020magnetic,xin2014electronic}. The in-plane FM interactions follow the Goodenough-Kanamori rule for Ni-O-Ni superexchange pathways, where the Ni-O-Ni bond angle is approximately 90$^\circ$ \cite{goodenough1955theory,goodenough1967theory,kanamori1959superexchange}. In contrast, the out-of-plane AFM interactions follow a Ni-O-O-Ni interaction path, which is weaker compared to the in-plane interactions \cite{goodenough1967theory,dey2021magnetostructural}. 

Previous studies on NiTiO$_{3}$ reported field-dependent thermal expansion and magnetostrictive effects, measured via dilatometry and neutron diffraction \cite{dey2021magnetostructural, harada2016magnetodielectric, dey2020magnetic}. These measurements show an anomaly at T$_N$ mostly associated with magneto-elastic coupling. Magnetic and phonon entropies obtained from heat capacity and dielectric data indicated the presence of spin-phonon interactions.

 In this study we performed elastic and inelastic neutron scattering measurements on powder and single crystals. NiTiO$_{3}$ has a magnon spectrum with a maximum energy cutoff of approximately 3.7 meV. Magnons show strong damping with increasing temperature while the inelastic intensity persist up to twice T${_N}$ due to short-range magnetic fluctuations persisting above the magnetic transition. The magnetic order parameter was used to calculate the critical exponent yielding a value of $\beta$ = 0.213(2), which indicates an anisotropic Heisenberg behavior. Spin-wave calculations yielded a Heisenberg-like Hamiltonian with six exchange interactions (Js). From the single-crystal inelastic neutron scattering, two magnon bands were discerned: a quasi-flat optic-like band and a dispersive acoustic-like band. An anisotropy gap of about 0.95 meV is observed at the Brillouin zone (BZ) center, which can be explained by single-ion anisotropy. The calculated frustration parameter of $\simeq$ 2 indicates that the system is weakly frustrated.

\section{Experiment and simulations}
\label{Experimental}

The NiTiO$_{3}$ sample was prepared using solid-state reaction method. NiO and TiO$_{2}$ powders were thoroughly mixed in a 1:1 molar ratio and reacted at 1350$^{\circ}$C in a tube furnace under argon atmosphere for 48 hours. Powder samples obtained from the solid-state reaction were used as the starting material for the single-crystal growth. The powder was ground using a mechanical mortar and pestle for 24 hours to achieve fine consistency. The powder was then pressed into long, thin rods (approximately 15 cm in length and 1 cm in diameter) using a hydrostatic press. The rods were subsequently annealed at 1350$^{\circ}$C for 12 hours in a tube furnace. Large single crystals were grown in the floating zone furnace. The resulting crystal, measuring 10 cm in length and 8 mm in diameter, was cut into two parts. The quality of the single crystal was confirmed by Laue diffraction, reflective light polarization, and single crystal X-ray diffraction. Magnetic susceptibility measurements were performed using a Physical Property Measurement System (PPMS).

The powder neutron scattering experiments were carried out using the VISION spectrometer at the Spallation Neutron Source (SNS) and the HB-2A powder diffractometer at the High Flux Isotope Reactor (HFIR) of Oak Ridge National Laboratory. A 10 g of powder sample was used for the VISION experiment, where elastic and inelastic data were acquired from 5 to 300 K, with a fixed final neutron energy of 3.5 meV. Data were collected over a finite momentum ($Q$) window: detector positions correspond to the low-$Q$ (LQ) path at 45$^{\circ}$ and the high-$Q$ (HQ) path at 135$^{\circ}$. For the HB-2A measurements, a 10 g powder sample was loaded into a vanadium can, and experiments were performed at 5, 200, 295, 400, 500, and 700 K using a Ge(115) monochromator and a neutron beam of 1.54 {\AA} wavelength. To ensure consistent temperature control within the HB-2A chamber, the sample environment was maintained under vacuum above 295 K and with helium exchange gas below 295 K. Refinement of the diffraction data confirmed the presence of a single NiTiO$_{3}$ phase at 5 K.

Single crystal neutron scattering measurements were performed using the Cold Neutron Triple Axis Spectrometer (CTAX) at HFIR. The first crystal weighing 8 g was aligned along the (HHL) plane, while a second crystal weighing 5 g, was aligned along the (H0L) plane. For the measurements, the final neutron energy was fixed at 4.8 meV and the energy resolution at the elastic line is 0.25 meV. A cooled Be filter was placed after the sample to reduce higher-order neutron contaminations. The magnon dispersions were simulated using SpinW and SU(N)NY ~\cite{toth2015linear,githubGitHubSunnySuiteSunnyjl}.

\begin{figure}[b]
\begin{center}
\includegraphics[width=8.6cm]{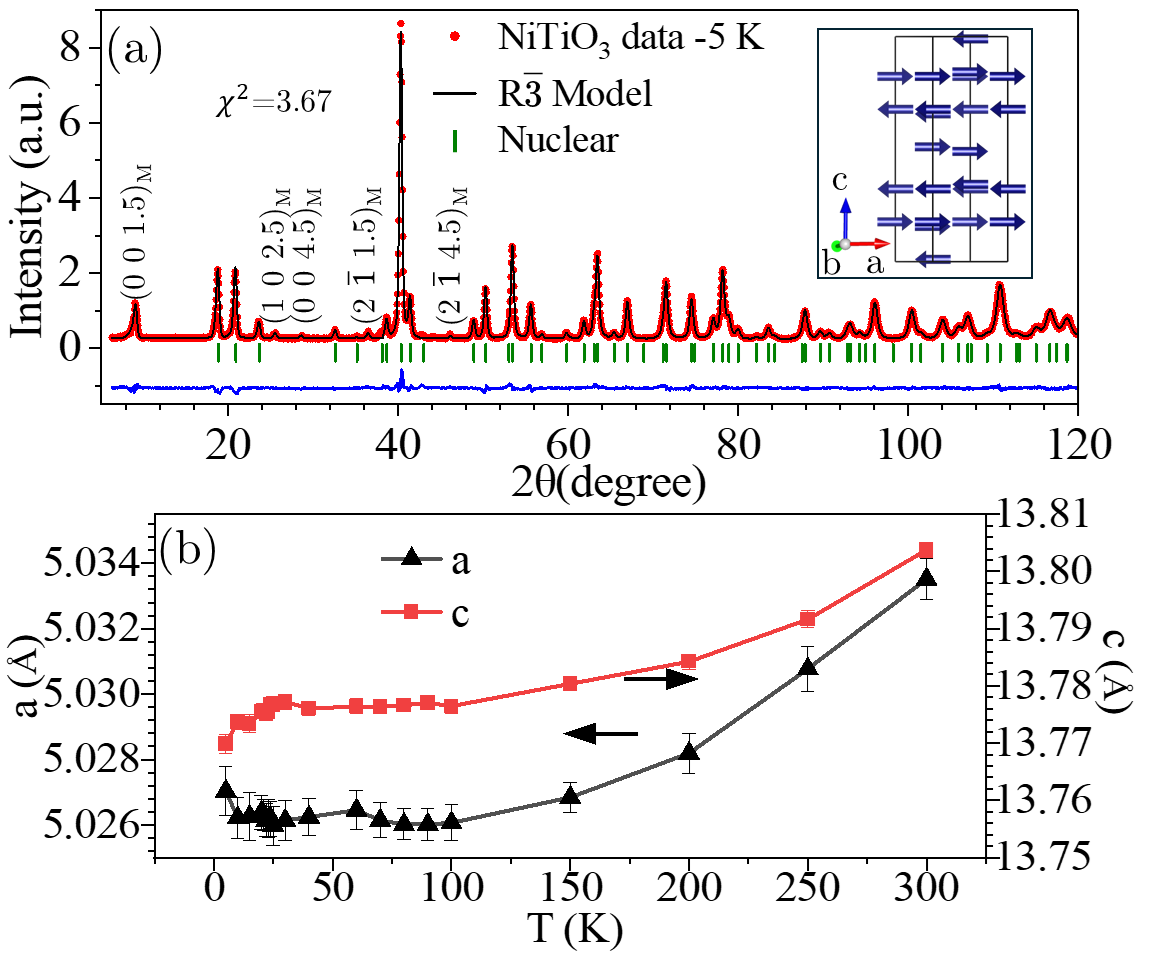}	
\end{center}
\caption{(a)Comparison of HB-2A diffraction data (red) with the fitting (black) at 5 K; the blue line corresponds to the difference between the data and the fitting. The green tick marks represent the nuclear Bragg peak positions and magnetic Braggs peaks are labeled. The inset is the A-type spin structure of the NiTiO$_3$. (b) Temperature dependence of the lattice constants $a$ and $c$ obtained from the VISION elastic data refinement. Error bars for the lattice constant $c$ are smaller than the size of the data points.} 
\label{fig2}
\end{figure}

\section{Results and Discussion}
\label{Results and Discussion}


The neutron powder diffraction measurements carried out at HB-2A did not show a structural transition in the 5 - 700 K temperature range. The HB-2A instrument can access $Q$ values as low as 0.45 Å$^{-1}$ with the selected wavelength for these measurements. Fig. 2(a) is a plot of analyzed neutron scattering data at 5 K using the Rietveld refinement. Analysis was performed using FullProf and SARAh-Representational Analysis to determine the nuclear and magnetic structure \cite{rodriguez1990fullprof,wills2000new}. The refinement of room-temperature data confirms the high quality of the sample, which consists solely of the NiTiO$_3$ phase with an R$\overline{3}$ space group and no detectable impurity phases. In the diffraction pattern, nuclear peaks are marked by green ticks, while magnetic peaks are labeled. The main (0 0 3/2) magnetic Bragg peak appears at 0.6725 Å$^{-1}$ (2$\theta$ = 9.05$^{\circ}$). The analysis confirmed that the  magnetic $\textit{k}$-vector for this system is (0, 0, 3/2). First, the calculation was performed allowing the magnetic moment to have an in-plane component ($M_X$). Refinement using both the in-plane ($M_X$) and out-of-plane ($M_Z$) components resulted in a better fit to certain magnetic peaks compared to refinement using only the in-plane component and the results of the two refinements are shown in Supplementary Fig. S1~\cite{sup}. The results shows a possible small out-of-plane magnetic moment, with a resultant in-plane magnetic moment per Ni atom of $M$ = 1.245(1) $\mu$$_B$. The magnetic space group is P$_S$$\overline{1}$  (\#2.7), consistent with previous reports ~\cite{dey2020magnetic}. The resulting magnetic structure is shown in the inset of Fig. 2(a). This space group allows moments both in the plane and out of the plane. The direction of the spin in the ab-plane was chosen to be along the $a$--axis, however, the exact direction can not be determined due to lack of directional information in the powder data.

Fig. 2(b) is a plot of the lattice constants $a$ and $c$ obtained from Rietveld refinement, based on the elastic neutron scattering data from VISION from 5 to 300 K. AFM magnetic Bragg peaks appear below the magnetic transition, $T$$_N$ $\approx$ 23 K. Above 70 K, $a$ and $c$ gradually increase with temperature. Below that, the $c$-lattice constant shows minimal change until 23 K, after which it drops. A negative thermal expansion (NTE) below 50 K has been previously reported for the $a$ lattice constant; however, our data does not support this observation. The drop at 23 K of the lattice parameters supports the magneto-elastic coupling previously demonstrated through dilatometry and heat capacity measurements~\cite{dey2020magnetic,harada2016magnetodielectric}.

\begin{figure}[t]
\centering 
\includegraphics[width=0.47\textwidth]{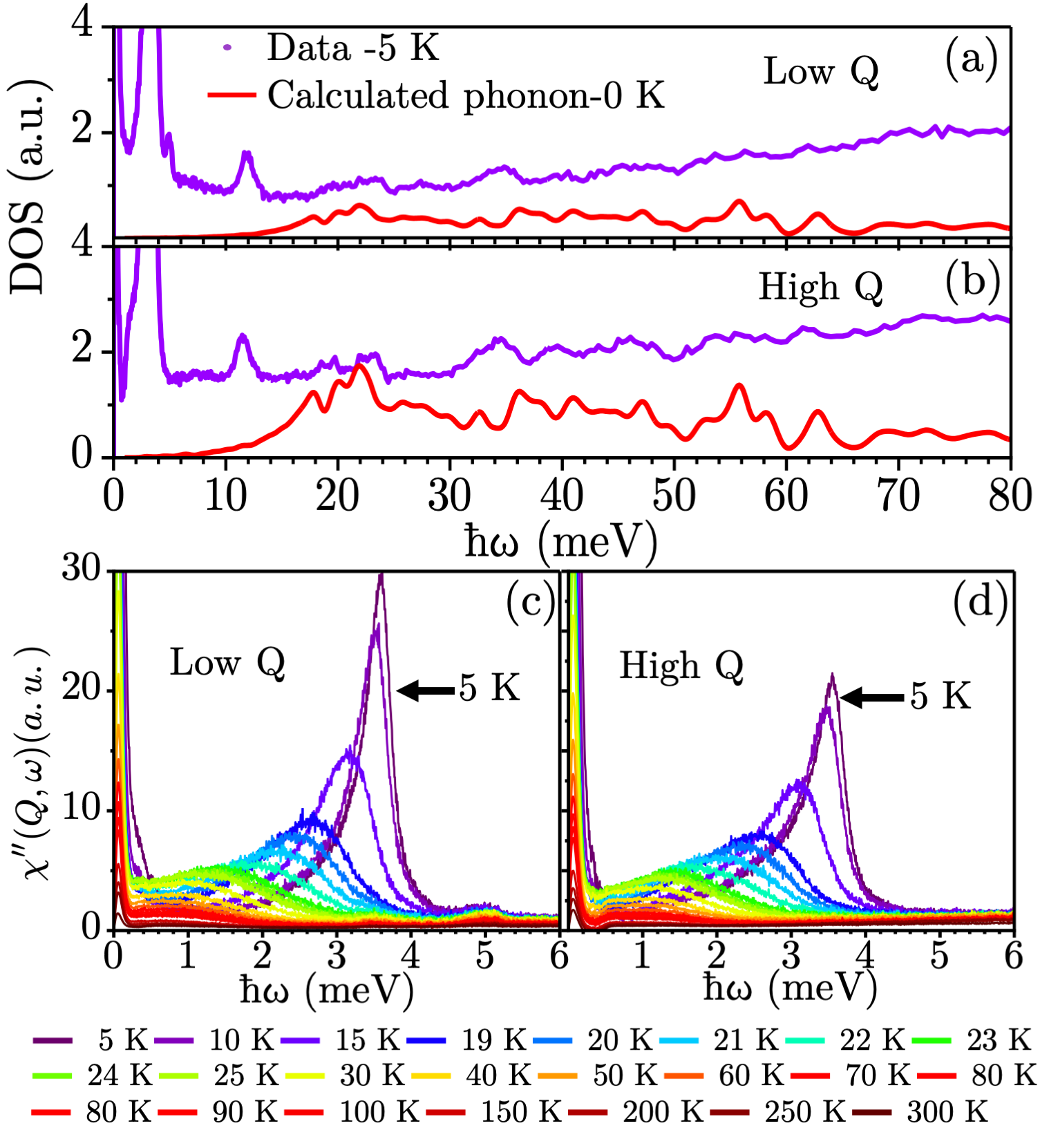}
\caption{The density of states (DOS) of the (a) LQ and (b) HQ trajectories at 5 K as a function of energy transfer ($\hbar$$\omega$), are compared to the calculated phonon DOS at 0 K. The $\chi''$(Q,$\omega$) ranging from 5 to 300 K for the (c) LQ trajectory and the (d) HQ trajectory. Bose correction and background subtraction were performed for the data at all temperatures.} 
\label{fig3}
\end{figure}


The magnon density of states (DOS) along the LQ and HQ trajectories at 5 K are compared to the calculated phonon DOS at 0 K (Figs. 3(a) and (b)). The neutron data contains both magnon and phonon contributions where the LQ trajectory has a higher magnetic contribution, whereas the HQ trajectory has less magnon and more phonon contribution due to the magnetic form factor of Ni$^{2+}$ and the fact that the HQ path reaches higher Q transfers. This distinction helps in identifying the phonon contribution in the data. A comparison of the DOS in Figs. 3(a) and (b) indicates that the phonon contribution is negligible below 10 meV. Note also that the prominent peak that is seen at ~12 meV is an instrument artifact due to $\lambda$/2 higher-order reflections from the analyzer.

\begin{figure}[b]
\centering 
\includegraphics[width=0.47\textwidth]{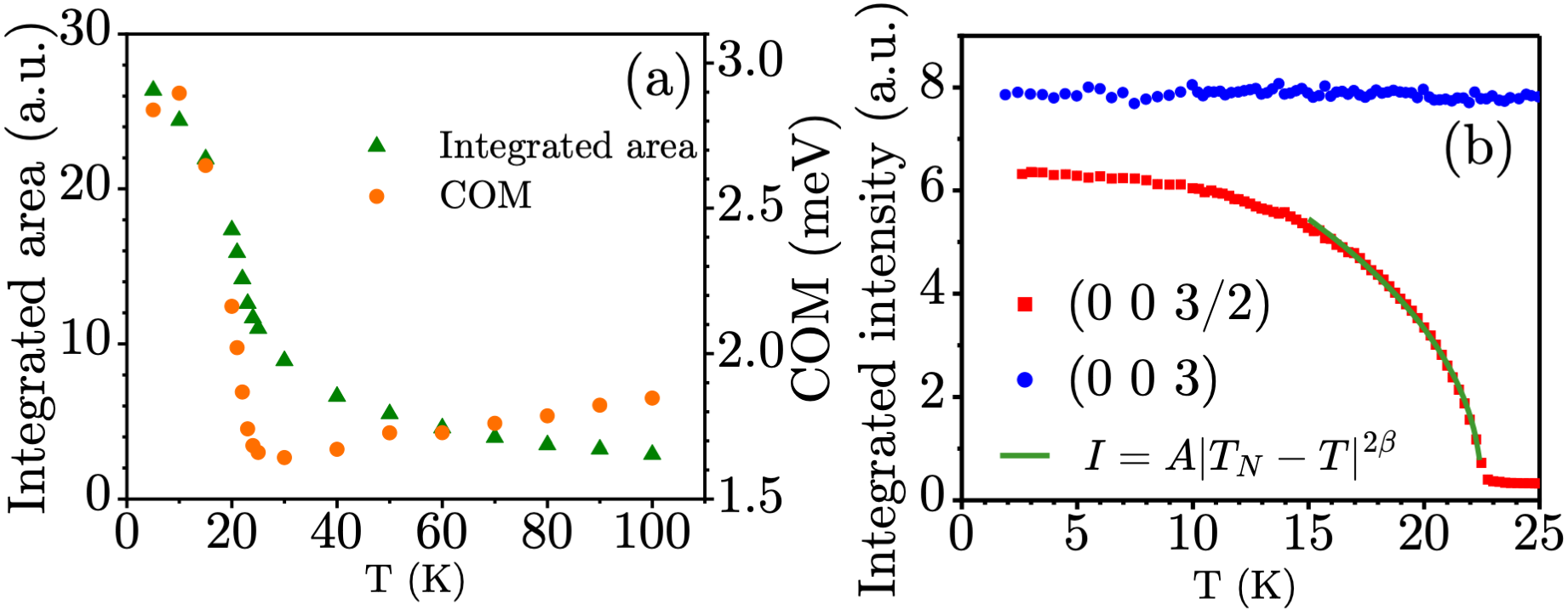}
\caption{(a) Integrated intensity(green) and the center of mass-COM(orange) of the inelastic peak in $\chi''$(Q,$\omega$) - $\hbar$$\omega$ plots as a function of temperature. The trapezoidal method used from 0.2 to 4.5 mev for the calculations and the error bars are smaller than the size of the data points. (b) Integrated intensity of the (0 0 3) nuclear peak(blue) and the (0 0 1.5) magnetic peak(red) as a function of temperature. The integrated intensity data for the magnetic peak is fitted using the equation provided in the inset, with the fitted line shown in green. The error bars are smaller than the size of the data points.} 
\label{fig4}
\end{figure}

Shown in Figs. 3(c) and (d) are plots of the temperature dependence of the $\chi''$(Q,$\omega$) along the LQ and HQ trajectories from 5 to 300 K. Unlike what was previously observed in CoTiO${_3}$, only a single sharp peak is present with a maximum magnon energy of about 4 meV. A small peak can be seen at 5 meV in the low-Q data which most likely is an instrument artifact previously observed in other measurements on VISION \cite{tao2023investigating}. The temperature changes of the magnon intensity are significant between 5 K and T$_N$ as seen in Figs. 3(c) and (d). The integrated area under the central peak and its center of mass (COM) are plotted as a function of temperature and shown in Fig. 4(a). It can be seen that, the area under the peak decreases sharply with T${_N}$ and shifts down in energy due to softening. Above T$_N$, the intensity persists up to 50 K, well above T$_N$. The persistence of inelastic intensity above T$_N$ supports the existence of short-range magnetic correlations at temperatures as high as 2T$_N$. Similar behavior has been reported in other AFM honeycomb systems such as CrCl$_3$ and CoTiO$_3$, where strong in-plane FM correlations above T$_N$ are responsible for short range ordering~\cite{mcguire2017magnetic,rathnayaka2024temperature}. Furthermore, the sharp change in the COM with increasing temperature reflects the sharp softening of the magnon energy as the system warms up. Thermal fluctuations and spin disorder contribute to the magnons damping.

\begin{figure}[t] 
\centering 
\includegraphics[width=0.47\textwidth]{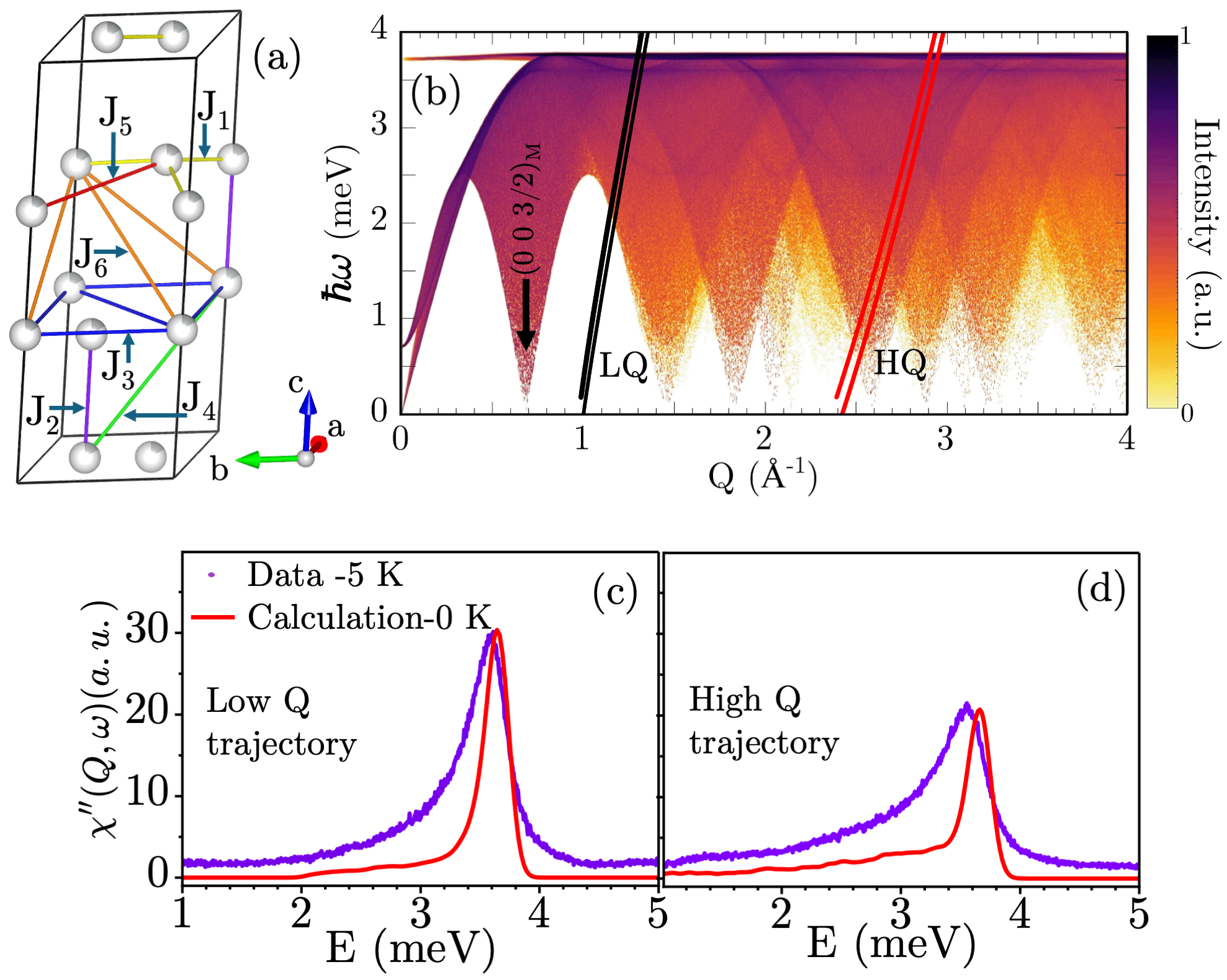}
\caption{(a) The six nearest neighbor exchange interactions, Js, between Ni atoms as listed in Table I. (b) Powder-averaged spin wave simulation \(S_{\bot}\left( Q,\omega \right)\) at 0 K. The (0, 0, 3/2) magnetic Bragg peak at 0.7 \AA${^{-1}}$ is labeled. The black lines correspond to the LQ trajectory, while the red lines represents the HQ path of the VISION instrument. Comparison of the scattering intensity obtained from the data at 5 K (purple) with the calculated \(S_{\bot}\left( Q,\omega \right)\) (red) at 0 K for the (c) LQ and (d) HQ trajectories, respectively.  } 
\label{fig5}
\end{figure}

A comparison of the integrated intensity of the (0, 0, 3) nuclear Bragg peak with the (0, 0, 3/2) magnetic peak is shown in Fig. 4(b). The data were obtained from the single crystal neutron scattering measurement. The nuclear peak intensity remains unchanged within the temperature range shown while the magnetic peak intensity decreases, as expected, and eventually vanishing at 22.5(1) K, which corresponds to $T$$_N$. The behavior of the magnetic peak can be described by Landau's theory of phase transition using, $I$=A($T$$_{N}$-$T$)$^{2\beta}$, with $\beta$ as the critical exponent \cite{landau1980statistical}. The critical exponent $\beta$, is found to be 0.213(2) from fitting the intensity. This value falls between a 3D Heisenberg model ($\beta$ = 0.368) and a 2D Ising model ($\beta$ = 0.125) \cite{landau1980statistical,gkebara2021determination,ahlberg2012critical} and is close to reported values for MnTiO$_3$, MnPS$_3$ and CrI$_3$, which are layered and highly anisotropic Heisenberg systems\cite{hwang2021spin,wildes2006static,liu2018three}.

\begin{table}[b]
\caption{Exchange interaction constants and anisotropy value used in the Heisenberg spin Hamiltonian for NiTiO$_3$, with the corresponding distances between Ni atoms (d) corresponding to the Js depicted in Fig. 5(a) and the number of bonds corresponding to the each exchange interaction (n) for an atom.}
\centering
\begin{tabular}{c c c c c c c c}
\hline\hline
 & $J$$_1$ & $J$$_2$ & $J$$_3$& $J$$_4$ & $J$$_5$ & $J$$_6$ & $D$ \\ [0.5ex] 
\hline
strength (meV)&-0.12  &0.24  &-0.08  &0.25  &-0.07  &0.23  & 0.12\\
d(\AA) &2.94  &4.12  & 5.03 &5.43 &5.83  &5.92 & -\\ 
 n &3  &1 &6  &6  &3  &3  & -\\
[1ex]
\hline
\end{tabular}
\label{table:nonlin}
\end{table}


In Figs. 5(a) and (b), the $\chi''$ is compared data at 5 K with the calculated spectrum at 0 K along the HQ and LQ paths of the VISION instrument. 
The calculated HQ and LQ $\chi''(Q,\omega)$ were obtained from modeling the magnon dispersions using linear spin wave theory and the powder-averaged to yield the magnon DOS. The interaction pairs used for the exchange model are shown in Fig.~5(c). The Hamiltonian for the linear spin wave calculation is: \[{H} = \sum_{\langle i, j \rangle}  J \left( S_i^x S_j^x + S_i^y S_j^y + S_i^z S_j^z \right) + D \sum_i \left(S_i^z\right)^2\]

where J is the exchange interaction constant, and $D$ is the single-ion anisotropy (SIA) constant. For the calculation, we used $S = 1$ and $\textit{k}$ = (0, 0, 3/2), along with the magnetic form factor of Ni$^{2+}$. Starting with nearest-neighbor in-plane and out-of-plane interactions, six exchange constants ($J_1$–$J_6$) were included in the spin Hamiltonian. $J_1$, $J_3$, and $J_5$ correspond to in-plane nearest-neighbor FM interactions, while $J_2$, $J_4$, and $J_6$ represent out-of-plane nearest-neighbor AFM interactions. The weighted residual factor was calculated by introducing different numbers of J's into the system, as detailed in the Supplementary section II~\cite{sup}. The six J
values listed in Table I yielded the lowest weighted residual factor, and the corresponding calculation is shown in supplementary figure S3 (f).

Fig.~5(b) shows the powder-averaged magnon dispersion calculated at 0 K. The DOS was obtained by integrating the intensities along the LQ and HQ paths, as shown in the figure. The simulation accounted for the energy resolution of the instrument, with the $Q$ resolution set to 0.2~\AA$^{-1}$. The model reproduces key features of the experimental data as shown in Figs. 5(c) and (d), including the maximum in $\chi''(Q,\omega)$ and the energy dependence of the DOS. However, the experimental data exhibit a broader distribution than the simulation, likely due to several factors. At low temperatures, dynamical interactions between spin waves, such as magnon-magnon interactions, are not included in linear spin wave theory, potentially leading to additional broadening of the magnon peaks~\cite{dyson1956thermodynamic}. Additionally, the calculations assume 0 K, while the experimental data were collected at 5 K. This small temperature difference could allow for thermal spin fluctuations, further contributing to peak broadening. This effect may be particularly relevant given that $J_1$ is extremely small ($\sim 1$ K) and $J_2$ is on the order of 3 K.

\begin{figure}[t]
\centering 
\includegraphics[width=0.47\textwidth]{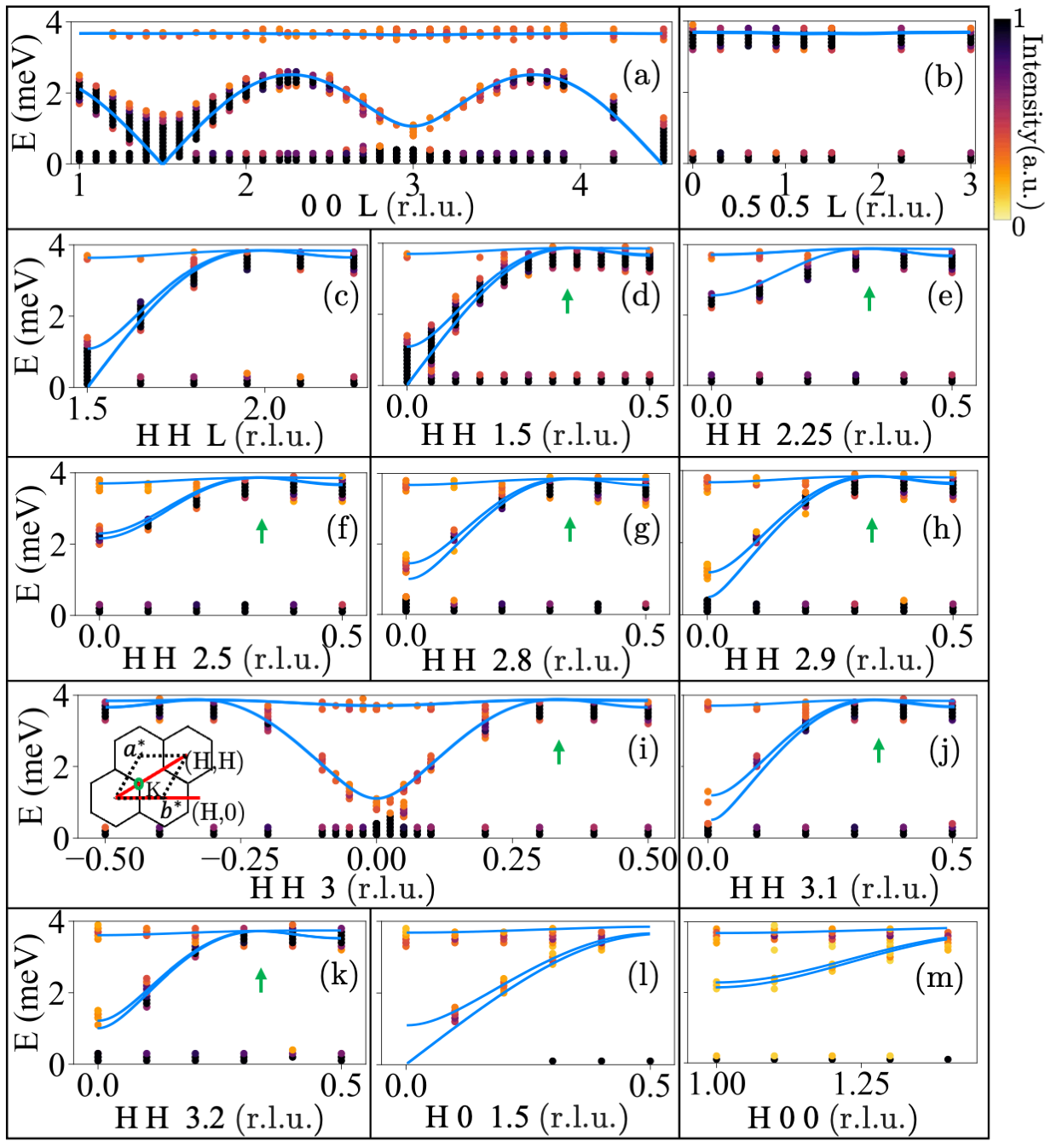}
\caption{Spin wave spectra of NiTiO$_3$ along the following directions:  (a) 0 0 L, (b) 0.5 0.5 L, (c) H H L from (0 0 1.5) to (0.5 0.5 2.25), (d) H H 1.5 (e) H H 2.25, (f) H H 2.5, (g) H H 2.8, (h) H H 2.9, (i) H H 3, (j) H H 3.1, (k) H H 3.2, (l) H 0 1.5, and (m) H 0 0, observed at 1.7 K with a fixed final energy of 4.8 meV. The solid dots represent the experimental data from the CTAX and the intensity scale is shown on top right with arbitrary units. The measurements were performed for constant Q cuts.  The blue line in each panel corresponds to the calculated spin wave dispersion at 0 K. The inset of 6(i) shows the honeycomb reciprocal space with measured (H,H) and (H,0) in-plane directions in the red line. The green circle is the K point of the system and the corresponding point in the color maps is marked by a green arrow.}
\label{fig6}
\end{figure}

Magnon dispersions along several high-symmetry directions were collected on CTAX using the single crystal (Fig.6).  Well-defined magnetic excitations are evident in the data, attributed to the good quality of the sample. Generally, honeycomb AFM systems exhibit four magnetic dispersion bands \cite{liu2022spin,choe2024magnetoelastic,huang2022topological,brehm2024magnon}. The four bands correspond to four magnon modes associated with the two magnetic sublattices from the two magnetic atoms, with each sublattice contributing to in-phase and out-of-phase modes. The data presented in Fig. 6 exhibits a quasi-flat optic-like band at approximately 3.7 meV and a low-energy acoustic-like dispersive band below 3.7 meV. The bands are most likely degenerate and are not resolved within the resolution of CTAX. Fig. 6(a) shows a dispersion along the direction perpendicular to the layers, indicative of a stronger inter-layer coupling which is consistent with the values of J${_2}$. The plots in Figs. 6(d)-(k) illustrate that at the high-symmetry K point, the lower dispersive band touches the high-energy quasi flat band.

The spin wave calculations were performed using the J parameters listed in Table I, and the calculated dispersions are directly compared with the experimental data, as shown in Fig. 6. The 0.95 meV gap observed at the (0, 0, 3) BZ center in Figs. 6(a) and 6(i) is the SIA along the z-axis, with the anisotropy value estimated to be approximately 0.12 meV. In the out-of-plane directions, as shown in Figs. 6(a) and 6(b), the calculations show only two bands, while along other directions, three bands are visible in the calculation. Without the anisotropy, only the flat band and a single low-energy dispersive band is present in the calculation; however, the presence of the anisotropy lifts the degeneracy of the lower dispersive band, resulting in two distinct bands, while the flat band remains unaffected. The band arising from the anisotropy is only present in the in-plane measurements because the intensity of the band resulted from anisotropy decreases near the (0, 0, 3/2) magnetic Bragg point and is not observed in out-of-plane measurements. The lifted degeneracy from the anisotropy cannot be seen in the data due to the limited resolution of the measurement.

Quasi flat magnon bands are characterized by highly localized spin waves resulting from strongly correlated spins, often exhibiting zero group velocity. Similar bands observed in layered kagome and triangular lattices where the spins have geometric induced frustration associated with DMI and SIA ~\cite{chisnell2015topological,riberolles2024chiral,seifert2022phase}. The bands in the NiTiO$_3$ should not be due to DMI because the collinear spin structure does not support DMI. The honeycomb lattice is not generally considered frustrated, but frustration can arise due to competing nearest-neighbor and next-nearest-neighbor magnetic interactions among hexagonally arranged magnetic ions~\cite{regnault1990phase,smirnova2009synthesis,seibel2013structure}. The frustration parameter ($f = |\Theta_{CW}|/T_N$) is an indicator of the degree of frustration in a system. Here the $\Theta_{CW}$ is the Curie-Weiss temperature(= -43.48 $\pm$ 0.32 K) of the NiTiO$_3$ sample and the corresponding calculations are shown in Supplementary section III~\cite{sup}.  For an ideal, non-frustrated system, $f$ would be close to 1. Several honeycomb systems with small $f$ values, such as BaNi$_2$X$_2$O$_8$ (X = V, P, As) and CrPSe$_3$ exhibit weak frustration~\cite{klyushina2017investigation,baithi2023incommensurate}. For NiTiO$_3$, the calculated $f \approx 2$, indicates weak frustration and this frustration could arise from competition between two channels of exchange coupling such as from the AFM interlayer and FM intra-layer coupling. Although there is no frustration for the in-plane FM interaction, long-range 2-dimensional FM order is not allowed due to Mermin-Wagner theorem \cite{giawei}. Therefore, the magnetic phase transition has to be a 3-dimensional order, and the establishment of a 3D order through inter-layer coupling is weakly frustrated, which could lead to the $f \approx 2$. 

 A recent paper on NiTiO$_3$  provided with exchange interaction constants closer to our model~\cite{kikuchi2025dirac}. However, in their model, nearest neighbor inter-plane interaction (J$_2$) was absent which is surprising given that this is a layered AFM system. Also, Ref.~\cite{kikuchi2025dirac} claims that Dirac magnon band crossing is observed without confirming the topological nature of the crossing. It is equally likely that the bands are approaching but not actually touching or crossing. Even if they cross, it does not mean that the crossing is topological. In our data we cannot confirm the presence of Dirac-like magnon band crossing. The dispersion is quite different from CoTiO${_3}$ where Dirac magnon crossing points were suggested. In the present case, accidental band touching at high symmetry K points is possible and does not imply Dirac crossing.

\section{Conclusion}
\label{Conclusion}

In conclusion, from neutron scattering measurements on single crystal and powder samples, observed a highly anisotropic spin state in NiTiO${_3}$. Unlike other ATiO${_3}$ ilmenites, the exchange interactions in NiTiO${_3}$ indicate relatively stronger inter-plane interactions compared to in-plane interactions ($J$$_1$,$J$$_3$,$J$$_5$ $<$ $J$$_2$,$J$$_4$,$J$$_6$), consistent with previous models of strong $J$$_{AFM}$ and weak $J$$_{FM}$ in Refs. \cite{goodenough1967theory,dey2020magnetic}. However, among the ATiO${_3}$ family, NiTiO${_3}$ exhibits the weakest overall exchange interactions. The spin Hamiltonian is Heisenberg in nature, with SIA along the z-axis. It is possible that the frustration arises from competing nearest-neighbor FM interactions within the honeycomb layer. The weak in-plane Js, comparable in magnitude to the SIA of the system, that results in a complex spin arrangement and contributes to the observed frustration. In contrast, other ATiO${_3}$ ilmenites do not exhibit such frustration, likely due to the anisotropies presents in the systems are not comparable with exchange interactions as in NiTiO${_3}$.

 NiTiO$_{3}$ presents a case where competing interactions prevent the system from settling into a simple unfrustrated FM ground state because in-plane J$_{1}$ is comparable to out-of-plane J$_{2}$. Thus it is possible that J$_{1}$ nearest neighbor is competing with J$_{2}$, next nearest neighbor. If J$_{2}$ is strong, then it can disrupt the FM order by promoting configurations that are AFM. In this system, the SIA is very weak and allows the spins to align in-plane. This is because the energy gain from the exchange interaction dominates over the penalty associated with aligning against the anisotropy.

\section{Acknowledgements}
\label{Acknowledgements}

The authors would like to acknowledge discussions with Yang Yang and Gia-wei Chern. They also thank Armando di Biase for assistance with the measurements. The work at the University of Virginia is supported by the Department of Energy, Grant number DE-FG02-01ER45927. This research used resources at the High Flux Isotope Reactor and Spallation Neutron Source,  both DOE Office of Science User Facilities operated by the Oak Ridge National Laboratory. 
The beam time was allocated to VISION on proposal number IPTS-31947.1, HB-2A on proposal number IPTS-31669.1, CTAX on proposal number IPTS-32124.1.



\bibliography{bibliography}

\begin{thebibliography}{66}%
\makeatletter
\providecommand \@ifxundefined [1]{%
 \@ifx{#1\undefined}
}%
\providecommand \@ifnum [1]{%
 \ifnum #1\expandafter \@firstoftwo
 \else \expandafter \@secondoftwo
 \fi
}%
\providecommand \@ifx [1]{%
 \ifx #1\expandafter \@firstoftwo
 \else \expandafter \@secondoftwo
 \fi
}%
\providecommand \natexlab [1]{#1}%
\providecommand \enquote  [1]{``#1''}%
\providecommand \bibnamefont  [1]{#1}%
\providecommand \bibfnamefont [1]{#1}%
\providecommand \citenamefont [1]{#1}%
\providecommand \href@noop [0]{\@secondoftwo}%
\providecommand \href [0]{\begingroup \@sanitize@url \@href}%
\providecommand \@href[1]{\@@startlink{#1}\@@href}%
\providecommand \@@href[1]{\endgroup#1\@@endlink}%
\providecommand \@sanitize@url [0]{\catcode `\\12\catcode `\$12\catcode `\&12\catcode `\#12\catcode `\^12\catcode `\_12\catcode `\%12\relax}%
\providecommand \@@startlink[1]{}%
\providecommand \@@endlink[0]{}%
\providecommand \url  [0]{\begingroup\@sanitize@url \@url }%
\providecommand \@url [1]{\endgroup\@href {#1}{\urlprefix }}%
\providecommand \urlprefix  [0]{URL }%
\providecommand \Eprint [0]{\href }%
\providecommand \doibase [0]{http://dx.doi.org/}%
\providecommand \selectlanguage [0]{\@gobble}%
\providecommand \bibinfo  [0]{\@secondoftwo}%
\providecommand \bibfield  [0]{\@secondoftwo}%
\providecommand \translation [1]{[#1]}%
\providecommand \BibitemOpen [0]{}%
\providecommand \bibitemStop [0]{}%
\providecommand \bibitemNoStop [0]{.\EOS\space}%
\providecommand \EOS [0]{\spacefactor3000\relax}%
\providecommand \BibitemShut  [1]{\csname bibitem#1\endcsname}%
\let\auto@bib@innerbib\@empty
\bibitem [{\citenamefont {Soumyanarayanan}\ \emph {et~al.}(2016)\citenamefont {Soumyanarayanan}, \citenamefont {Reyren}, \citenamefont {Fert},\ and\ \citenamefont {Panagopoulos}}]{soumyanarayanan2016emergent}%
  \BibitemOpen
  \bibfield  {author} {\bibinfo {author} {\bibfnamefont {A.}~\bibnamefont {Soumyanarayanan}}, \bibinfo {author} {\bibfnamefont {N.}~\bibnamefont {Reyren}}, \bibinfo {author} {\bibfnamefont {A.}~\bibnamefont {Fert}}, \ and\ \bibinfo {author} {\bibfnamefont {C.}~\bibnamefont {Panagopoulos}},\ }\bibfield  {title} {\enquote {\bibinfo {title} {Emergent phenomena induced by spin--orbit coupling at surfaces and interfaces},}\ }\href@noop {} {\bibfield  {journal} {\bibinfo  {journal} {Nature}\ }\textbf {\bibinfo {volume} {539}},\ \bibinfo {pages} {509--517} (\bibinfo {year} {2016})}\BibitemShut {NoStop}%
\bibitem [{\citenamefont {Miao}\ \emph {et~al.}(2018)\citenamefont {Miao}, \citenamefont {Xu}, \citenamefont {Zhu}, \citenamefont {Zhou},\ and\ \citenamefont {Sun}}]{miao20182d}%
  \BibitemOpen
  \bibfield  {author} {\bibinfo {author} {\bibfnamefont {N.}~\bibnamefont {Miao}}, \bibinfo {author} {\bibfnamefont {B.}~\bibnamefont {Xu}}, \bibinfo {author} {\bibfnamefont {L.}~\bibnamefont {Zhu}}, \bibinfo {author} {\bibfnamefont {J.}~\bibnamefont {Zhou}}, \ and\ \bibinfo {author} {\bibfnamefont {Z.}~\bibnamefont {Sun}},\ }\bibfield  {title} {\enquote {\bibinfo {title} {2{D} {I}ntrinsic {F}erromagnets from van der {W}aals {A}ntiferromagnets},}\ }\href@noop {} {\bibfield  {journal} {\bibinfo  {journal} {Journal of the American Chemical Society}\ }\textbf {\bibinfo {volume} {140}},\ \bibinfo {pages} {2417--2420} (\bibinfo {year} {2018})}\BibitemShut {NoStop}%
\bibitem [{\citenamefont {Fert}(2008)}]{fert2008nobel}%
  \BibitemOpen
  \bibfield  {author} {\bibinfo {author} {\bibfnamefont {A.}~\bibnamefont {Fert}},\ }\bibfield  {title} {\enquote {\bibinfo {title} {Nobel {L}ecture: {O}rigin, development, and future of spintronics},}\ }\href@noop {} {\bibfield  {journal} {\bibinfo  {journal} {Reviews of modern physics}\ }\textbf {\bibinfo {volume} {80}},\ \bibinfo {pages} {1517--1530} (\bibinfo {year} {2008})}\BibitemShut {NoStop}%
\bibitem [{\citenamefont {Wolf}\ \emph {et~al.}(2001)\citenamefont {Wolf}, \citenamefont {Awschalom}, \citenamefont {Buhrman}, \citenamefont {Daughton}, \citenamefont {von Moln{\'a}r}, \citenamefont {Roukes}, \citenamefont {Chtchelkanova},\ and\ \citenamefont {Treger}}]{wolf2001spintronics}%
  \BibitemOpen
  \bibfield  {author} {\bibinfo {author} {\bibfnamefont {S.~A.}\ \bibnamefont {Wolf}}, \bibinfo {author} {\bibfnamefont {D.~D.}\ \bibnamefont {Awschalom}}, \bibinfo {author} {\bibfnamefont {R.~A.}\ \bibnamefont {Buhrman}}, \bibinfo {author} {\bibfnamefont {J.~M.}\ \bibnamefont {Daughton}}, \bibinfo {author} {\bibfnamefont {von~S.}\ \bibnamefont {von Moln{\'a}r}}, \bibinfo {author} {\bibfnamefont {M.~L.}\ \bibnamefont {Roukes}}, \bibinfo {author} {\bibfnamefont {A.~Yu}\ \bibnamefont {Chtchelkanova}}, \ and\ \bibinfo {author} {\bibfnamefont {D.~M.}\ \bibnamefont {Treger}},\ }\bibfield  {title} {\enquote {\bibinfo {title} {Spintronics: a spin-based electronics vision for the future},}\ }\href@noop {} {\bibfield  {journal} {\bibinfo  {journal} {science}\ }\textbf {\bibinfo {volume} {294}},\ \bibinfo {pages} {1488--1495} (\bibinfo {year} {2001})}\BibitemShut {NoStop}%
\bibitem [{\citenamefont {Felser}\ \emph {et~al.}(2007)\citenamefont {Felser}, \citenamefont {Fecher},\ and\ \citenamefont {Balke}}]{felser2007spintronics}%
  \BibitemOpen
  \bibfield  {author} {\bibinfo {author} {\bibfnamefont {C.}~\bibnamefont {Felser}}, \bibinfo {author} {\bibfnamefont {G.~H.}\ \bibnamefont {Fecher}}, \ and\ \bibinfo {author} {\bibfnamefont {B.}~\bibnamefont {Balke}},\ }\bibfield  {title} {\enquote {\bibinfo {title} {Spintronics: a challenge for materials science and solid-state chemistry},}\ }\href@noop {} {\bibfield  {journal} {\bibinfo  {journal} {Angewandte Chemie International Edition}\ }\textbf {\bibinfo {volume} {46}},\ \bibinfo {pages} {668--699} (\bibinfo {year} {2007})}\BibitemShut {NoStop}%
\bibitem [{\citenamefont {Tokura}\ \emph {et~al.}(2019)\citenamefont {Tokura}, \citenamefont {Yasuda},\ and\ \citenamefont {Tsukazaki}}]{tokura2019magnetic}%
  \BibitemOpen
  \bibfield  {author} {\bibinfo {author} {\bibfnamefont {Y.}~\bibnamefont {Tokura}}, \bibinfo {author} {\bibfnamefont {K.}~\bibnamefont {Yasuda}}, \ and\ \bibinfo {author} {\bibfnamefont {A.}~\bibnamefont {Tsukazaki}},\ }\bibfield  {title} {\enquote {\bibinfo {title} {Magnetic topological insulators},}\ }\href@noop {} {\bibfield  {journal} {\bibinfo  {journal} {Nature Reviews Physics}\ }\textbf {\bibinfo {volume} {1}},\ \bibinfo {pages} {126--143} (\bibinfo {year} {2019})}\BibitemShut {NoStop}%
\bibitem [{\citenamefont {{\v{S}}mejkal}\ \emph {et~al.}(2018)\citenamefont {{\v{S}}mejkal}, \citenamefont {Mokrousov}, \citenamefont {Yan},\ and\ \citenamefont {MacDonald}}]{vsmejkal2018topological}%
  \BibitemOpen
  \bibfield  {author} {\bibinfo {author} {\bibfnamefont {L.}~\bibnamefont {{\v{S}}mejkal}}, \bibinfo {author} {\bibfnamefont {Y.}~\bibnamefont {Mokrousov}}, \bibinfo {author} {\bibfnamefont {B.}~\bibnamefont {Yan}}, \ and\ \bibinfo {author} {\bibfnamefont {A.~H.}\ \bibnamefont {MacDonald}},\ }\bibfield  {title} {\enquote {\bibinfo {title} {Topological antiferromagnetic spintronics},}\ }\href@noop {} {\bibfield  {journal} {\bibinfo  {journal} {Nature physics}\ }\textbf {\bibinfo {volume} {14}},\ \bibinfo {pages} {242--251} (\bibinfo {year} {2018})}\BibitemShut {NoStop}%
\bibitem [{\citenamefont {Bishop}\ \emph {et~al.}(2012)\citenamefont {Bishop}, \citenamefont {Li}, \citenamefont {Farnell},\ and\ \citenamefont {Campbell}}]{bishop2012frustrated}%
  \BibitemOpen
  \bibfield  {author} {\bibinfo {author} {\bibfnamefont {R.~F.}\ \bibnamefont {Bishop}}, \bibinfo {author} {\bibfnamefont {P.~H.~Y.}\ \bibnamefont {Li}}, \bibinfo {author} {\bibfnamefont {Damian J.~J.}\ \bibnamefont {Farnell}}, \ and\ \bibinfo {author} {\bibfnamefont {C.~E.}\ \bibnamefont {Campbell}},\ }\bibfield  {title} {\enquote {\bibinfo {title} {The frustrated {H}eisenberg antiferromagnet on the honeycomb lattice: {J}1--{J}2 model},}\ }\href@noop {} {\bibfield  {journal} {\bibinfo  {journal} {Journal of Physics: Condensed Matter}\ }\textbf {\bibinfo {volume} {24}},\ \bibinfo {pages} {236002} (\bibinfo {year} {2012})}\BibitemShut {NoStop}%
\bibitem [{\citenamefont {Hu}\ and\ \citenamefont {Kan}(2019)}]{hu2019progress}%
  \BibitemOpen
  \bibfield  {author} {\bibinfo {author} {\bibfnamefont {T.}~\bibnamefont {Hu}}\ and\ \bibinfo {author} {\bibfnamefont {E.}~\bibnamefont {Kan}},\ }\bibfield  {title} {\enquote {\bibinfo {title} {Progress and prospects in low-dimensional multiferroic materials},}\ }\href@noop {} {\bibfield  {journal} {\bibinfo  {journal} {Wiley Interdisciplinary Reviews: Computational Molecular Science}\ }\textbf {\bibinfo {volume} {9}},\ \bibinfo {pages} {e1409} (\bibinfo {year} {2019})}\BibitemShut {NoStop}%
\bibitem [{\citenamefont {Dey}\ \emph {et~al.}(2021)\citenamefont {Dey}, \citenamefont {Sauerland}, \citenamefont {Ouladdiaf}, \citenamefont {Beauvois}, \citenamefont {Wadepohl},\ and\ \citenamefont {Klingeler}}]{dey2021magnetostructural}%
  \BibitemOpen
  \bibfield  {author} {\bibinfo {author} {\bibfnamefont {K.}~\bibnamefont {Dey}}, \bibinfo {author} {\bibfnamefont {S.}~\bibnamefont {Sauerland}}, \bibinfo {author} {\bibfnamefont {B.}~\bibnamefont {Ouladdiaf}}, \bibinfo {author} {\bibfnamefont {K.}~\bibnamefont {Beauvois}}, \bibinfo {author} {\bibfnamefont {H.}~\bibnamefont {Wadepohl}}, \ and\ \bibinfo {author} {\bibfnamefont {R.}~\bibnamefont {Klingeler}},\ }\bibfield  {title} {\enquote {\bibinfo {title} {Magnetostructural coupling in ilmenite-type {NiTiO}$_3$},}\ }\href@noop {} {\bibfield  {journal} {\bibinfo  {journal} {Physical Review B}\ }\textbf {\bibinfo {volume} {103}},\ \bibinfo {pages} {134438} (\bibinfo {year} {2021})}\BibitemShut {NoStop}%
\bibitem [{\citenamefont {Schneeloch}\ \emph {et~al.}(2022)\citenamefont {Schneeloch}, \citenamefont {Tao}, \citenamefont {Cheng}, \citenamefont {Daemen}, \citenamefont {Xu}, \citenamefont {Zhang},\ and\ \citenamefont {Louca}}]{schneeloch2022gapless}%
  \BibitemOpen
  \bibfield  {author} {\bibinfo {author} {\bibfnamefont {J.~A.}\ \bibnamefont {Schneeloch}}, \bibinfo {author} {\bibfnamefont {Y.}~\bibnamefont {Tao}}, \bibinfo {author} {\bibfnamefont {Y.}~\bibnamefont {Cheng}}, \bibinfo {author} {\bibfnamefont {L.}~\bibnamefont {Daemen}}, \bibinfo {author} {\bibfnamefont {G.}~\bibnamefont {Xu}}, \bibinfo {author} {\bibfnamefont {Q.}~\bibnamefont {Zhang}}, \ and\ \bibinfo {author} {\bibfnamefont {D.}~\bibnamefont {Louca}},\ }\bibfield  {title} {\enquote {\bibinfo {title} {Gapless {D}irac magnons in {C}r{C}l$_3$},}\ }\href@noop {} {\bibfield  {journal} {\bibinfo  {journal} {npj Quantum Materials}\ }\textbf {\bibinfo {volume} {7}},\ \bibinfo {pages} {66} (\bibinfo {year} {2022})}\BibitemShut {NoStop}%
\bibitem [{\citenamefont {Chen}\ \emph {et~al.}(2023)\citenamefont {Chen}, \citenamefont {Huang},\ and\ \citenamefont {Fu}}]{chen2023damped}%
  \BibitemOpen
  \bibfield  {author} {\bibinfo {author} {\bibfnamefont {Q.~H.}\ \bibnamefont {Chen}}, \bibinfo {author} {\bibfnamefont {F.~J.}\ \bibnamefont {Huang}}, \ and\ \bibinfo {author} {\bibfnamefont {Y.~P.}\ \bibnamefont {Fu}},\ }\bibfield  {title} {\enquote {\bibinfo {title} {Damped topological magnons in honeycomb antiferromagnets},}\ }\href@noop {} {\bibfield  {journal} {\bibinfo  {journal} {Physical Review B}\ }\textbf {\bibinfo {volume} {108}},\ \bibinfo {pages} {024409} (\bibinfo {year} {2023})}\BibitemShut {NoStop}%
\bibitem [{\citenamefont {Ederer}\ and\ \citenamefont {Fennie}(2008)}]{ederer2008electric}%
  \BibitemOpen
  \bibfield  {author} {\bibinfo {author} {\bibfnamefont {C.}~\bibnamefont {Ederer}}\ and\ \bibinfo {author} {\bibfnamefont {C.~J.}\ \bibnamefont {Fennie}},\ }\bibfield  {title} {\enquote {\bibinfo {title} {Electric-field switchable magnetization via the {D}zyaloshinskii--{M}oriya interaction: {F}e{T}i{O}$_3$ versus {B}i{F}e{O}$_3$},}\ }\href@noop {} {\bibfield  {journal} {\bibinfo  {journal} {Journal of Physics: Condensed Matter}\ }\textbf {\bibinfo {volume} {20}},\ \bibinfo {pages} {434219} (\bibinfo {year} {2008})}\BibitemShut {NoStop}%
\bibitem [{\citenamefont {Niu}\ \emph {et~al.}(2017)\citenamefont {Niu}, \citenamefont {Hanke}, \citenamefont {Buhl}, \citenamefont {Bihlmayer}, \citenamefont {Wortmann}, \citenamefont {Bl{\"u}gel},\ and\ \citenamefont {Mokrousov}}]{niu2017quantum}%
  \BibitemOpen
  \bibfield  {author} {\bibinfo {author} {\bibfnamefont {C.}~\bibnamefont {Niu}}, \bibinfo {author} {\bibfnamefont {J.~P.}\ \bibnamefont {Hanke}}, \bibinfo {author} {\bibfnamefont {P.~M.}\ \bibnamefont {Buhl}}, \bibinfo {author} {\bibfnamefont {G.}~\bibnamefont {Bihlmayer}}, \bibinfo {author} {\bibfnamefont {D.}~\bibnamefont {Wortmann}}, \bibinfo {author} {\bibfnamefont {S.}~\bibnamefont {Bl{\"u}gel}}, \ and\ \bibinfo {author} {\bibfnamefont {Y.}~\bibnamefont {Mokrousov}},\ }\bibfield  {title} {\enquote {\bibinfo {title} {Quantum spin hall effect and topological phase transitions in honeycomb antiferromagnets},}\ }\href@noop {} {\bibfield  {journal} {\bibinfo  {journal} {arXiv preprint arXiv:1705.07035}\ } (\bibinfo {year} {2017})}\BibitemShut {NoStop}%
\bibitem [{\citenamefont {Haldane}(1988)}]{haldane1988model}%
  \BibitemOpen
  \bibfield  {author} {\bibinfo {author} {\bibfnamefont {F.~Duncan~M.}\ \bibnamefont {Haldane}},\ }\bibfield  {title} {\enquote {\bibinfo {title} {Model for a quantum hall effect without {L}andau levels: Condensed-matter realization of the" parity anomaly"},}\ }\href@noop {} {\bibfield  {journal} {\bibinfo  {journal} {Physical review letters}\ }\textbf {\bibinfo {volume} {61}},\ \bibinfo {pages} {2015} (\bibinfo {year} {1988})}\BibitemShut {NoStop}%
\bibitem [{\citenamefont {Go}\ \emph {et~al.}(2024)\citenamefont {Go}, \citenamefont {An}, \citenamefont {Lee},\ and\ \citenamefont {Kim}}]{go2024magnon}%
  \BibitemOpen
  \bibfield  {author} {\bibinfo {author} {\bibfnamefont {G.}~\bibnamefont {Go}}, \bibinfo {author} {\bibfnamefont {D.}~\bibnamefont {An}}, \bibinfo {author} {\bibfnamefont {H.~W.}\ \bibnamefont {Lee}}, \ and\ \bibinfo {author} {\bibfnamefont {S.~K.}\ \bibnamefont {Kim}},\ }\bibfield  {title} {\enquote {\bibinfo {title} {Magnon orbital {N}ernst effect in honeycomb antiferromagnets without spin--orbit coupling},}\ }\href@noop {} {\bibfield  {journal} {\bibinfo  {journal} {Nano Letters}\ }\textbf {\bibinfo {volume} {24}},\ \bibinfo {pages} {5968--5974} (\bibinfo {year} {2024})}\BibitemShut {NoStop}%
\bibitem [{\citenamefont {McClarty}(2022)}]{mcclarty2022topological}%
  \BibitemOpen
  \bibfield  {author} {\bibinfo {author} {\bibfnamefont {P.~A.}\ \bibnamefont {McClarty}},\ }\bibfield  {title} {\enquote {\bibinfo {title} {Topological magnons: A review},}\ }\href@noop {} {\bibfield  {journal} {\bibinfo  {journal} {Annual Review of Condensed Matter Physics}\ }\textbf {\bibinfo {volume} {13}},\ \bibinfo {pages} {171--190} (\bibinfo {year} {2022})}\BibitemShut {NoStop}%
\bibitem [{\citenamefont {Yuan}\ \emph {et~al.}(2020{\natexlab{a}})\citenamefont {Yuan}, \citenamefont {Khait}, \citenamefont {Shu}, \citenamefont {Chou}, \citenamefont {Stone}, \citenamefont {Clancy}, \citenamefont {Paramekanti},\ and\ \citenamefont {Kim}}]{yuan2020dirac}%
  \BibitemOpen
  \bibfield  {author} {\bibinfo {author} {\bibfnamefont {B.}~\bibnamefont {Yuan}}, \bibinfo {author} {\bibfnamefont {I.}~\bibnamefont {Khait}}, \bibinfo {author} {\bibfnamefont {G.~J.}\ \bibnamefont {Shu}}, \bibinfo {author} {\bibfnamefont {F.~C.}\ \bibnamefont {Chou}}, \bibinfo {author} {\bibfnamefont {M.~B.}\ \bibnamefont {Stone}}, \bibinfo {author} {\bibfnamefont {J.~P.}\ \bibnamefont {Clancy}}, \bibinfo {author} {\bibfnamefont {A.}~\bibnamefont {Paramekanti}}, \ and\ \bibinfo {author} {\bibfnamefont {Y.~J.}\ \bibnamefont {Kim}},\ }\bibfield  {title} {\enquote {\bibinfo {title} {Dirac magnons in a honeycomb lattice quantum {XY} magnet {CoTiO}$_3$},}\ }\href@noop {} {\bibfield  {journal} {\bibinfo  {journal} {Physical Review X}\ }\textbf {\bibinfo {volume} {10}},\ \bibinfo {pages} {011062} (\bibinfo {year} {2020}{\natexlab{a}})}\BibitemShut {NoStop}%
\bibitem [{\citenamefont {Schneeloch}\ \emph {et~al.}(2024)\citenamefont {Schneeloch}, \citenamefont {Daemen},\ and\ \citenamefont {Louca}}]{schneeloch2024antiferromagnetic}%
  \BibitemOpen
  \bibfield  {author} {\bibinfo {author} {\bibfnamefont {J.~A.}\ \bibnamefont {Schneeloch}}, \bibinfo {author} {\bibfnamefont {L.}~\bibnamefont {Daemen}}, \ and\ \bibinfo {author} {\bibfnamefont {D.}~\bibnamefont {Louca}},\ }\bibfield  {title} {\enquote {\bibinfo {title} {Antiferromagnetic-ferromagnetic homostructures with {D}irac magnons in the van der waals magnet {C}r{I}$_3$},}\ }\href@noop {} {\bibfield  {journal} {\bibinfo  {journal} {Physical Review B}\ }\textbf {\bibinfo {volume} {109}},\ \bibinfo {pages} {024409} (\bibinfo {year} {2024})}\BibitemShut {NoStop}%
\bibitem [{\citenamefont {Rathnayaka}\ \emph {et~al.}(2024)\citenamefont {Rathnayaka}, \citenamefont {Daemen}, \citenamefont {Schneeloch}, \citenamefont {Cheng},\ and\ \citenamefont {Louca}}]{rathnayaka2024temperature}%
  \BibitemOpen
  \bibfield  {author} {\bibinfo {author} {\bibfnamefont {S.}~\bibnamefont {Rathnayaka}}, \bibinfo {author} {\bibfnamefont {L.}~\bibnamefont {Daemen}}, \bibinfo {author} {\bibfnamefont {J.~A.}\ \bibnamefont {Schneeloch}}, \bibinfo {author} {\bibfnamefont {Y.}~\bibnamefont {Cheng}}, \ and\ \bibinfo {author} {\bibfnamefont {D.}~\bibnamefont {Louca}},\ }\bibfield  {title} {\enquote {\bibinfo {title} {Temperature dependence of magnetic excitations in the topological insulator {CoTiO}$_3$},}\ }\href@noop {} {\bibfield  {journal} {\bibinfo  {journal} {Physical Review B}\ }\textbf {\bibinfo {volume} {109}},\ \bibinfo {pages} {174432} (\bibinfo {year} {2024})}\BibitemShut {NoStop}%
\bibitem [{\citenamefont {Goodenough}\ and\ \citenamefont {Stickler}(1967)}]{goodenough1967theory}%
  \BibitemOpen
  \bibfield  {author} {\bibinfo {author} {\bibfnamefont {J.~B.}\ \bibnamefont {Goodenough}}\ and\ \bibinfo {author} {\bibfnamefont {J.~J.}\ \bibnamefont {Stickler}},\ }\bibfield  {title} {\enquote {\bibinfo {title} {Theory of the magnetic properties of the ilmenites {MTiO}$_3$},}\ }\href@noop {} {\bibfield  {journal} {\bibinfo  {journal} {Physical Review}\ }\textbf {\bibinfo {volume} {164}},\ \bibinfo {pages} {768} (\bibinfo {year} {1967})}\BibitemShut {NoStop}%
\bibitem [{\citenamefont {Osmond}(1964)}]{osmond1964magnetic}%
  \BibitemOpen
  \bibfield  {author} {\bibinfo {author} {\bibfnamefont {W.~P.}\ \bibnamefont {Osmond}},\ }\bibfield  {title} {\enquote {\bibinfo {title} {Magnetic exchange interactions in ilmenites' {MeTiO}$_3$ ({Me= Mn, Fe, Co and Ni})},}\ }\href@noop {} {\bibfield  {journal} {\bibinfo  {journal} {British Journal of Applied Physics}\ }\textbf {\bibinfo {volume} {15}},\ \bibinfo {pages} {1377} (\bibinfo {year} {1964})}\BibitemShut {NoStop}%
\bibitem [{\citenamefont {Shirane}\ \emph {et~al.}(1959)\citenamefont {Shirane}, \citenamefont {SJ},\ and\ \citenamefont {Ishikawa}}]{shirane1959neutron}%
  \BibitemOpen
  \bibfield  {author} {\bibinfo {author} {\bibfnamefont {G.}~\bibnamefont {Shirane}}, \bibinfo {author} {\bibfnamefont {P.}~\bibnamefont {SJ}}, \ and\ \bibinfo {author} {\bibfnamefont {Y.}~\bibnamefont {Ishikawa}},\ }\bibfield  {title} {\enquote {\bibinfo {title} {Neutron diffraction study of antiferromagnetic {MnTiO}$_3$ and {NiTiO}$_3$},}\ }\href@noop {} {\bibfield  {journal} {\bibinfo  {journal} {Journal of the Physical Society of Japan}\ }\textbf {\bibinfo {volume} {14}},\ \bibinfo {pages} {1352--1360} (\bibinfo {year} {1959})}\BibitemShut {NoStop}%
\bibitem [{\citenamefont {Kato}\ \emph {et~al.}(1983)\citenamefont {Kato}, \citenamefont {Funahashi}, \citenamefont {Yamada},\ and\ \citenamefont {Takei}}]{kato1983coexistence}%
  \BibitemOpen
  \bibfield  {author} {\bibinfo {author} {\bibfnamefont {H.}~\bibnamefont {Kato}}, \bibinfo {author} {\bibfnamefont {S.}~\bibnamefont {Funahashi}}, \bibinfo {author} {\bibfnamefont {M.}~\bibnamefont {Yamada}}, \ and\ \bibinfo {author} {\bibfnamefont {H.}~\bibnamefont {Takei}},\ }\bibfield  {title} {\enquote {\bibinfo {title} {Coexistence of ising-and xy-like magnons in {FeTiO}$_3$},}\ }\href@noop {} {\bibfield  {journal} {\bibinfo  {journal} {Journal of Magnetism and Magnetic Materials}\ }\textbf {\bibinfo {volume} {31}},\ \bibinfo {pages} {617--618} (\bibinfo {year} {1983})}\BibitemShut {NoStop}%
\bibitem [{\citenamefont {Hwang}\ \emph {et~al.}(2021)\citenamefont {Hwang}, \citenamefont {Lee}, \citenamefont {Chung}, \citenamefont {Ikeuchi}, \citenamefont {Garlea}, \citenamefont {Yamauchi}, \citenamefont {Akatsu},\ and\ \citenamefont {Shamoto}}]{hwang2021spin}%
  \BibitemOpen
  \bibfield  {author} {\bibinfo {author} {\bibfnamefont {I.~Y.}\ \bibnamefont {Hwang}}, \bibinfo {author} {\bibfnamefont {K.~H.}\ \bibnamefont {Lee}}, \bibinfo {author} {\bibfnamefont {J.~H.}\ \bibnamefont {Chung}}, \bibinfo {author} {\bibfnamefont {K.}~\bibnamefont {Ikeuchi}}, \bibinfo {author} {\bibfnamefont {V.~O.}\ \bibnamefont {Garlea}}, \bibinfo {author} {\bibfnamefont {H.}~\bibnamefont {Yamauchi}}, \bibinfo {author} {\bibfnamefont {M.}~\bibnamefont {Akatsu}}, \ and\ \bibinfo {author} {\bibfnamefont {S.}~\bibnamefont {Shamoto}},\ }\bibfield  {title} {\enquote {\bibinfo {title} {Spin wave excitations in honeycomb antiferromagnet {MnTiO}$_3$},}\ }\href@noop {} {\bibfield  {journal} {\bibinfo  {journal} {Journal of the Physical Society of Japan}\ }\textbf {\bibinfo {volume} {90}},\ \bibinfo {pages} {064708} (\bibinfo {year} {2021})}\BibitemShut {NoStop}%
\bibitem [{\citenamefont {Yuan}\ \emph {et~al.}(2020{\natexlab{b}})\citenamefont {Yuan}, \citenamefont {Stone}, \citenamefont {Shu}, \citenamefont {Chou}, \citenamefont {Rao}, \citenamefont {Clancy},\ and\ \citenamefont {Kim}}]{yuan2020spin}%
  \BibitemOpen
  \bibfield  {author} {\bibinfo {author} {\bibfnamefont {B.}~\bibnamefont {Yuan}}, \bibinfo {author} {\bibfnamefont {M.~B.}\ \bibnamefont {Stone}}, \bibinfo {author} {\bibfnamefont {G.~J.}\ \bibnamefont {Shu}}, \bibinfo {author} {\bibfnamefont {F.~C.}\ \bibnamefont {Chou}}, \bibinfo {author} {\bibfnamefont {X.}~\bibnamefont {Rao}}, \bibinfo {author} {\bibfnamefont {J.~P.}\ \bibnamefont {Clancy}}, \ and\ \bibinfo {author} {\bibfnamefont {Y.~J.}\ \bibnamefont {Kim}},\ }\bibfield  {title} {\enquote {\bibinfo {title} {Spin-orbit exciton in a honeycomb lattice magnet {CoTiO}$_3$: Revealing a link between magnetism in $d$-and $f$-electron systems},}\ }\href@noop {} {\bibfield  {journal} {\bibinfo  {journal} {Physical Review B}\ }\textbf {\bibinfo {volume} {102}},\ \bibinfo {pages} {134404} (\bibinfo {year} {2020}{\natexlab{b}})}\BibitemShut {NoStop}%
\bibitem [{\citenamefont {Yuan}\ \emph {et~al.}(2024)\citenamefont {Yuan}, \citenamefont {Horsley}, \citenamefont {Stone}, \citenamefont {Butch}, \citenamefont {Xu}, \citenamefont {Shu}, \citenamefont {Clancy},\ and\ \citenamefont {Kim}}]{yuan2024field}%
  \BibitemOpen
  \bibfield  {author} {\bibinfo {author} {\bibfnamefont {B.}~\bibnamefont {Yuan}}, \bibinfo {author} {\bibfnamefont {E.}~\bibnamefont {Horsley}}, \bibinfo {author} {\bibfnamefont {M.~B.}\ \bibnamefont {Stone}}, \bibinfo {author} {\bibfnamefont {N.~P.}\ \bibnamefont {Butch}}, \bibinfo {author} {\bibfnamefont {G.}~\bibnamefont {Xu}}, \bibinfo {author} {\bibfnamefont {G.~J.}\ \bibnamefont {Shu}}, \bibinfo {author} {\bibfnamefont {J.~P.}\ \bibnamefont {Clancy}}, \ and\ \bibinfo {author} {\bibfnamefont {Y.~J.}\ \bibnamefont {Kim}},\ }\bibfield  {title} {\enquote {\bibinfo {title} {Field-dependent magnons in the honeycomb antiferromagnet {CoTiO}$_3$},}\ }\href@noop {} {\bibfield  {journal} {\bibinfo  {journal} {Physical Review B}\ }\textbf {\bibinfo {volume} {109}},\ \bibinfo {pages} {174440} (\bibinfo {year} {2024})}\BibitemShut {NoStop}%
\bibitem [{\citenamefont {Elliot}\ \emph {et~al.}(2021)\citenamefont {Elliot}, \citenamefont {McClarty}, \citenamefont {Prabhakaran}, \citenamefont {Johnson}, \citenamefont {Walker}, \citenamefont {Manuel},\ and\ \citenamefont {Coldea}}]{elliot2021order}%
  \BibitemOpen
  \bibfield  {author} {\bibinfo {author} {\bibfnamefont {M.}~\bibnamefont {Elliot}}, \bibinfo {author} {\bibfnamefont {P.~A.}\ \bibnamefont {McClarty}}, \bibinfo {author} {\bibfnamefont {D.}~\bibnamefont {Prabhakaran}}, \bibinfo {author} {\bibfnamefont {R.~D.}\ \bibnamefont {Johnson}}, \bibinfo {author} {\bibfnamefont {H.~C.}\ \bibnamefont {Walker}}, \bibinfo {author} {\bibfnamefont {P.}~\bibnamefont {Manuel}}, \ and\ \bibinfo {author} {\bibfnamefont {R.}~\bibnamefont {Coldea}},\ }\bibfield  {title} {\enquote {\bibinfo {title} {Order-by-disorder from bond-dependent exchange and intensity signature of nodal quasiparticles in a honeycomb cobaltate},}\ }\href@noop {} {\bibfield  {journal} {\bibinfo  {journal} {Nature Communications}\ }\textbf {\bibinfo {volume} {12}},\ \bibinfo {pages} {3936} (\bibinfo {year} {2021})}\BibitemShut {NoStop}%
\bibitem [{\citenamefont {Newnham}\ \emph {et~al.}(1964)\citenamefont {Newnham}, \citenamefont {Fang},\ and\ \citenamefont {Santoro}}]{newnham1964crystal}%
  \BibitemOpen
  \bibfield  {author} {\bibinfo {author} {\bibfnamefont {R.~E.}\ \bibnamefont {Newnham}}, \bibinfo {author} {\bibfnamefont {J.~H.}\ \bibnamefont {Fang}}, \ and\ \bibinfo {author} {\bibfnamefont {R.~P.}\ \bibnamefont {Santoro}},\ }\bibfield  {title} {\enquote {\bibinfo {title} {Crystal structure and magnetic properties of {CoTiO}$_3$},}\ }\href@noop {} {\bibfield  {journal} {\bibinfo  {journal} {Acta Crystallographica}\ }\textbf {\bibinfo {volume} {17}},\ \bibinfo {pages} {240--242} (\bibinfo {year} {1964})}\BibitemShut {NoStop}%
\bibitem [{\citenamefont {Yamaguchi}\ \emph {et~al.}(1986)\citenamefont {Yamaguchi}, \citenamefont {Kato}, \citenamefont {Takei}, \citenamefont {Goldman},\ and\ \citenamefont {Shirane}}]{yamaguchi1986re}%
  \BibitemOpen
  \bibfield  {author} {\bibinfo {author} {\bibfnamefont {Y.}~\bibnamefont {Yamaguchi}}, \bibinfo {author} {\bibfnamefont {H.}~\bibnamefont {Kato}}, \bibinfo {author} {\bibfnamefont {H.}~\bibnamefont {Takei}}, \bibinfo {author} {\bibfnamefont {A.~I.}\ \bibnamefont {Goldman}}, \ and\ \bibinfo {author} {\bibfnamefont {G.}~\bibnamefont {Shirane}},\ }\bibfield  {title} {\enquote {\bibinfo {title} {Re-examination of the magnetic structure of {FeTiO}$_3$},}\ }\href@noop {} {\bibfield  {journal} {\bibinfo  {journal} {Solid state communications}\ }\textbf {\bibinfo {volume} {59}},\ \bibinfo {pages} {865--868} (\bibinfo {year} {1986})}\BibitemShut {NoStop}%
\bibitem [{\citenamefont {Charilaou}\ \emph {et~al.}(2012)\citenamefont {Charilaou}, \citenamefont {Sheptyakov}, \citenamefont {L{\"o}ffler},\ and\ \citenamefont {Gehring}}]{charilaou2012large}%
  \BibitemOpen
  \bibfield  {author} {\bibinfo {author} {\bibfnamefont {M.}~\bibnamefont {Charilaou}}, \bibinfo {author} {\bibfnamefont {D.}~\bibnamefont {Sheptyakov}}, \bibinfo {author} {\bibfnamefont {J.~F.}\ \bibnamefont {L{\"o}ffler}}, \ and\ \bibinfo {author} {\bibfnamefont {A.~U.}\ \bibnamefont {Gehring}},\ }\bibfield  {title} {\enquote {\bibinfo {title} {Large spontaneous magnetostriction in {FeTiO}$_3$ and adjustable magnetic configuration in {Fe(III)}-doped {FeTiO}$_3$},}\ }\href@noop {} {\bibfield  {journal} {\bibinfo  {journal} {Physical Review B—Condensed Matter and Materials Physics}\ }\textbf {\bibinfo {volume} {86}},\ \bibinfo {pages} {024439} (\bibinfo {year} {2012})}\BibitemShut {NoStop}%
\bibitem [{\citenamefont {Acharya}\ and\ \citenamefont {Choudhary}(2015)}]{acharya2015structural}%
  \BibitemOpen
  \bibfield  {author} {\bibinfo {author} {\bibfnamefont {T.}~\bibnamefont {Acharya}}\ and\ \bibinfo {author} {\bibfnamefont {R.~N.~P.}\ \bibnamefont {Choudhary}},\ }\bibfield  {title} {\enquote {\bibinfo {title} {Structural, ferroelectric, and electrical properties of {N}i{T}i{O}$_3$ ceramic},}\ }\href@noop {} {\bibfield  {journal} {\bibinfo  {journal} {Journal of Electronic Materials}\ }\textbf {\bibinfo {volume} {44}},\ \bibinfo {pages} {271--280} (\bibinfo {year} {2015})}\BibitemShut {NoStop}%
\bibitem [{\citenamefont {Bamzai}\ \emph {et~al.}(2011)\citenamefont {Bamzai}, \citenamefont {Gupta}, \citenamefont {Kotru},\ and\ \citenamefont {Wanklyn}}]{bamzai2011dielectric}%
  \BibitemOpen
  \bibfield  {author} {\bibinfo {author} {\bibfnamefont {K.~K.}\ \bibnamefont {Bamzai}}, \bibinfo {author} {\bibfnamefont {V.}~\bibnamefont {Gupta}}, \bibinfo {author} {\bibfnamefont {P.~N.}\ \bibnamefont {Kotru}}, \ and\ \bibinfo {author} {\bibfnamefont {B.~M.}\ \bibnamefont {Wanklyn}},\ }\bibfield  {title} {\enquote {\bibinfo {title} {Dielectric and {AC} conductivity behaviour of flux grown nickel titanate ({N}i{T}i{O}$_3$) crystal},}\ }\href@noop {} {\bibfield  {journal} {\bibinfo  {journal} {Ferroelectrics}\ }\textbf {\bibinfo {volume} {413}},\ \bibinfo {pages} {328--341} (\bibinfo {year} {2011})}\BibitemShut {NoStop}%
\bibitem [{\citenamefont {Zhang}\ \emph {et~al.}(2015)\citenamefont {Zhang}, \citenamefont {Lu}, \citenamefont {Li}, \citenamefont {Fan}, \citenamefont {Liang},\ and\ \citenamefont {Han}}]{zhang2015structural}%
  \BibitemOpen
  \bibfield  {author} {\bibinfo {author} {\bibfnamefont {X.}~\bibnamefont {Zhang}}, \bibinfo {author} {\bibfnamefont {B.}~\bibnamefont {Lu}}, \bibinfo {author} {\bibfnamefont {R.}~\bibnamefont {Li}}, \bibinfo {author} {\bibfnamefont {C.}~\bibnamefont {Fan}}, \bibinfo {author} {\bibfnamefont {Z.}~\bibnamefont {Liang}}, \ and\ \bibinfo {author} {\bibfnamefont {P.}~\bibnamefont {Han}},\ }\bibfield  {title} {\enquote {\bibinfo {title} {Structural, electronic and optical properties of ilmenite {A}i{T}i{O}$_3$ ({A}= {F}e, {C}o, {N}i)},}\ }\href@noop {} {\bibfield  {journal} {\bibinfo  {journal} {Materials Science in Semiconductor Processing}\ }\textbf {\bibinfo {volume} {39}},\ \bibinfo {pages} {6--16} (\bibinfo {year} {2015})}\BibitemShut {NoStop}%
\bibitem [{\citenamefont {Dey}\ \emph {et~al.}(2020)\citenamefont {Dey}, \citenamefont {Sauerland}, \citenamefont {Werner}, \citenamefont {Skourski}, \citenamefont {Abdel-Hafiez}, \citenamefont {Bag}, \citenamefont {Singh},\ and\ \citenamefont {Klingeler}}]{dey2020magnetic}%
  \BibitemOpen
  \bibfield  {author} {\bibinfo {author} {\bibfnamefont {K.}~\bibnamefont {Dey}}, \bibinfo {author} {\bibfnamefont {S.}~\bibnamefont {Sauerland}}, \bibinfo {author} {\bibfnamefont {J.}~\bibnamefont {Werner}}, \bibinfo {author} {\bibfnamefont {Y.}~\bibnamefont {Skourski}}, \bibinfo {author} {\bibfnamefont {M.}~\bibnamefont {Abdel-Hafiez}}, \bibinfo {author} {\bibfnamefont {R.}~\bibnamefont {Bag}}, \bibinfo {author} {\bibfnamefont {S.}~\bibnamefont {Singh}}, \ and\ \bibinfo {author} {\bibfnamefont {R.}~\bibnamefont {Klingeler}},\ }\bibfield  {title} {\enquote {\bibinfo {title} {Magnetic phase diagram and magnetoelastic coupling of {NiTiO}$_3$},}\ }\href@noop {} {\bibfield  {journal} {\bibinfo  {journal} {Physical Review B}\ }\textbf {\bibinfo {volume} {101}},\ \bibinfo {pages} {195122} (\bibinfo {year} {2020})}\BibitemShut {NoStop}%
\bibitem [{\citenamefont {Xin}\ \emph {et~al.}(2014)\citenamefont {Xin}, \citenamefont {Wang}, \citenamefont {Sui}, \citenamefont {Wang}, \citenamefont {Wang}, \citenamefont {Zhao}, \citenamefont {Liu}, \citenamefont {Li},\ and\ \citenamefont {Liu}}]{xin2014electronic}%
  \BibitemOpen
  \bibfield  {author} {\bibinfo {author} {\bibfnamefont {C.}~\bibnamefont {Xin}}, \bibinfo {author} {\bibfnamefont {Y.}~\bibnamefont {Wang}}, \bibinfo {author} {\bibfnamefont {Y.}~\bibnamefont {Sui}}, \bibinfo {author} {\bibfnamefont {Y.}~\bibnamefont {Wang}}, \bibinfo {author} {\bibfnamefont {X.}~\bibnamefont {Wang}}, \bibinfo {author} {\bibfnamefont {K.}~\bibnamefont {Zhao}}, \bibinfo {author} {\bibfnamefont {Z.}~\bibnamefont {Liu}}, \bibinfo {author} {\bibfnamefont {B.}~\bibnamefont {Li}}, \ and\ \bibinfo {author} {\bibfnamefont {X.}~\bibnamefont {Liu}},\ }\bibfield  {title} {\enquote {\bibinfo {title} {Electronic, magnetic and multiferroic properties of magnetoelectric {NiTiO}$_3$},}\ }\href@noop {} {\bibfield  {journal} {\bibinfo  {journal} {Journal of alloys and compounds}\ }\textbf {\bibinfo {volume} {613}},\ \bibinfo {pages} {401--406} (\bibinfo {year} {2014})}\BibitemShut {NoStop}%
\bibitem [{\citenamefont {Goodenough}(1955)}]{goodenough1955theory}%
  \BibitemOpen
  \bibfield  {author} {\bibinfo {author} {\bibfnamefont {J.~B.}\ \bibnamefont {Goodenough}},\ }\bibfield  {title} {\enquote {\bibinfo {title} {Theory of the role of covalence in the perovskite-type manganites {[La,M(II)]MnO$_3$}},}\ }\href@noop {} {\bibfield  {journal} {\bibinfo  {journal} {Physical Review}\ }\textbf {\bibinfo {volume} {100}},\ \bibinfo {pages} {564} (\bibinfo {year} {1955})}\BibitemShut {NoStop}%
\bibitem [{\citenamefont {Kanamori}(1959)}]{kanamori1959superexchange}%
  \BibitemOpen
  \bibfield  {author} {\bibinfo {author} {\bibfnamefont {J.}~\bibnamefont {Kanamori}},\ }\bibfield  {title} {\enquote {\bibinfo {title} {Superexchange interaction and symmetry properties of electron orbitals},}\ }\href@noop {} {\bibfield  {journal} {\bibinfo  {journal} {Journal of Physics and Chemistry of Solids}\ }\textbf {\bibinfo {volume} {10}},\ \bibinfo {pages} {87--98} (\bibinfo {year} {1959})}\BibitemShut {NoStop}%
\bibitem [{\citenamefont {Harada}\ \emph {et~al.}(2016)\citenamefont {Harada}, \citenamefont {Balhorn}, \citenamefont {Hazi}, \citenamefont {Kemei},\ and\ \citenamefont {Seshadri}}]{harada2016magnetodielectric}%
  \BibitemOpen
  \bibfield  {author} {\bibinfo {author} {\bibfnamefont {J.~K.}\ \bibnamefont {Harada}}, \bibinfo {author} {\bibfnamefont {L.}~\bibnamefont {Balhorn}}, \bibinfo {author} {\bibfnamefont {J.}~\bibnamefont {Hazi}}, \bibinfo {author} {\bibfnamefont {M.~C.}\ \bibnamefont {Kemei}}, \ and\ \bibinfo {author} {\bibfnamefont {R.}~\bibnamefont {Seshadri}},\ }\bibfield  {title} {\enquote {\bibinfo {title} {Magnetodielectric coupling in the ilmenites {MTiO$_3$(M=Co,Ni)}},}\ }\href@noop {} {\bibfield  {journal} {\bibinfo  {journal} {Physical Review B}\ }\textbf {\bibinfo {volume} {93}},\ \bibinfo {pages} {104404} (\bibinfo {year} {2016})}\BibitemShut {NoStop}%
\bibitem [{\citenamefont {Toth}\ and\ \citenamefont {Lake}(2015)}]{toth2015linear}%
  \BibitemOpen
  \bibfield  {author} {\bibinfo {author} {\bibfnamefont {S.}~\bibnamefont {Toth}}\ and\ \bibinfo {author} {\bibfnamefont {B.}~\bibnamefont {Lake}},\ }\bibfield  {title} {\enquote {\bibinfo {title} {Linear spin wave theory for single-{Q} incommensurate magnetic structures},}\ }\href@noop {} {\bibfield  {journal} {\bibinfo  {journal} {Journal of Physics: Condensed Matter}\ }\textbf {\bibinfo {volume} {27}},\ \bibinfo {pages} {166002} (\bibinfo {year} {2015})}\BibitemShut {NoStop}%
\bibitem [{git()}]{githubGitHubSunnySuiteSunnyjl}%
  \BibitemOpen
  \href@noop {} {\enquote {\bibinfo {title} {{G}it{H}ub - {S}unny{S}uite/{S}unny.jl: {S}pin dynamics and generalization to {S}{U}({N}) coherent states --- github.com},}\ }\bibinfo {howpublished} {\url{https://github.com/sunnysuite/sunny.jl}}\BibitemShut {NoStop}%
\bibitem [{\citenamefont {Rodriguez-Carvajal}(1990)}]{rodriguez1990fullprof}%
  \BibitemOpen
  \bibfield  {author} {\bibinfo {author} {\bibfnamefont {J.}~\bibnamefont {Rodriguez-Carvajal}},\ }\bibfield  {title} {\enquote {\bibinfo {title} {Fullprof: a program for {R}ietveld refinement and pattern matching analysis},}\ }in\ \href@noop {} {\emph {\bibinfo {booktitle} {satellite meeting on powder diffraction of the XV congress of the IUCr}}},\ Vol.\ \bibinfo {volume} {127}\ (\bibinfo {organization} {Toulouse, France:[sn]},\ \bibinfo {year} {1990})\BibitemShut {NoStop}%
\bibitem [{\citenamefont {Wills}(2000)}]{wills2000new}%
  \BibitemOpen
  \bibfield  {author} {\bibinfo {author} {\bibfnamefont {A.~S.}\ \bibnamefont {Wills}},\ }\bibfield  {title} {\enquote {\bibinfo {title} {A new protocol for the determination of magnetic structures using simulated annealing and representational analysis {(SARAh)}},}\ }\href@noop {} {\bibfield  {journal} {\bibinfo  {journal} {Physica B: Condensed Matter}\ }\textbf {\bibinfo {volume} {276}},\ \bibinfo {pages} {680--681} (\bibinfo {year} {2000})}\BibitemShut {NoStop}%
\bibitem [{sup()}]{sup}%
  \BibitemOpen
  \bibfield  {title} {\enquote {\bibinfo {title} {See {S}upplemental {M}aterial [url] for the calculations of {R} for different {J} models},}\ }\href@noop {} {\ }\BibitemShut {NoStop}%
\bibitem [{\citenamefont {Tao}\ \emph {et~al.}(2023)\citenamefont {Tao}, \citenamefont {Daemen}, \citenamefont {Cheng}, \citenamefont {Neuefeind},\ and\ \citenamefont {Louca}}]{tao2023investigating}%
  \BibitemOpen
  \bibfield  {author} {\bibinfo {author} {\bibfnamefont {Y.}~\bibnamefont {Tao}}, \bibinfo {author} {\bibfnamefont {L.}~\bibnamefont {Daemen}}, \bibinfo {author} {\bibfnamefont {Y.}~\bibnamefont {Cheng}}, \bibinfo {author} {\bibfnamefont {J.~C.}\ \bibnamefont {Neuefeind}}, \ and\ \bibinfo {author} {\bibfnamefont {D.}~\bibnamefont {Louca}},\ }\bibfield  {title} {\enquote {\bibinfo {title} {Investigating the magnetoelastic properties in {FeSn} and {Fe$_3$Sn$_2$} flat band metals},}\ }\href@noop {} {\bibfield  {journal} {\bibinfo  {journal} {Physical Review B}\ }\textbf {\bibinfo {volume} {107}},\ \bibinfo {pages} {174407} (\bibinfo {year} {2023})}\BibitemShut {NoStop}%
\bibitem [{\citenamefont {McGuire}\ \emph {et~al.}(2017)\citenamefont {McGuire}, \citenamefont {Clark}, \citenamefont {Kc}, \citenamefont {Chance}, \citenamefont {Jellison~Jr}, \citenamefont {Cooper}, \citenamefont {Xu},\ and\ \citenamefont {Sales}}]{mcguire2017magnetic}%
  \BibitemOpen
  \bibfield  {author} {\bibinfo {author} {\bibfnamefont {M.~A.}\ \bibnamefont {McGuire}}, \bibinfo {author} {\bibfnamefont {G.}~\bibnamefont {Clark}}, \bibinfo {author} {\bibfnamefont {S.}~\bibnamefont {Kc}}, \bibinfo {author} {\bibfnamefont {W.~M.}\ \bibnamefont {Chance}}, \bibinfo {author} {\bibfnamefont {G.~E.}\ \bibnamefont {Jellison~Jr}}, \bibinfo {author} {\bibfnamefont {V.~R.}\ \bibnamefont {Cooper}}, \bibinfo {author} {\bibfnamefont {X.}~\bibnamefont {Xu}}, \ and\ \bibinfo {author} {\bibfnamefont {B.~C.}\ \bibnamefont {Sales}},\ }\bibfield  {title} {\enquote {\bibinfo {title} {Magnetic behavior and spin-lattice coupling in cleavable van der waals layered {C}r{C}l$_3$ crystals},}\ }\href@noop {} {\bibfield  {journal} {\bibinfo  {journal} {Physical Review Materials}\ }\textbf {\bibinfo {volume} {1}},\ \bibinfo {pages} {014001} (\bibinfo {year} {2017})}\BibitemShut {NoStop}%
\bibitem [{\citenamefont {Landau}\ \emph {et~al.}(1980)\citenamefont {Landau}, \citenamefont {Lifshitz},\ and\ \citenamefont {Pitaevskii}}]{landau1980statistical}%
  \BibitemOpen
  \bibfield  {author} {\bibinfo {author} {\bibfnamefont {L.~D.}\ \bibnamefont {Landau}}, \bibinfo {author} {\bibfnamefont {E.~M.}\ \bibnamefont {Lifshitz}}, \ and\ \bibinfo {author} {\bibfnamefont {L.~P.}\ \bibnamefont {Pitaevskii}},\ }\href@noop {} {\emph {\bibinfo {title} {Statistical physics: theory of the condensed state}}},\ Vol.~\bibinfo {volume} {9}\ (\bibinfo  {publisher} {Butterworth-Heinemann},\ \bibinfo {year} {1980})\BibitemShut {NoStop}%
\bibitem [{\citenamefont {Gebara}\ and\ \citenamefont {Hasiak}(2021)}]{gkebara2021determination}%
  \BibitemOpen
  \bibfield  {author} {\bibinfo {author} {\bibfnamefont {P.}~\bibnamefont {Gebara}}\ and\ \bibinfo {author} {\bibfnamefont {M.}~\bibnamefont {Hasiak}},\ }\bibfield  {title} {\enquote {\bibinfo {title} {Determination of phase transition and critical behavior of the as-cast {GdGeSi-(x) type alloys (where x= Ni,Nd and Pr)}},}\ }\href@noop {} {\bibfield  {journal} {\bibinfo  {journal} {Materials}\ }\textbf {\bibinfo {volume} {14}},\ \bibinfo {pages} {185} (\bibinfo {year} {2021})}\BibitemShut {NoStop}%
\bibitem [{\citenamefont {Ahlberg}(2012)}]{ahlberg2012critical}%
  \BibitemOpen
  \bibfield  {author} {\bibinfo {author} {\bibfnamefont {M.}~\bibnamefont {Ahlberg}},\ }\emph {\bibinfo {title} {Critical Phenomena and Exchange Coupling in Magnetic Heterostructures}},\ \href@noop {} {Ph.D. thesis},\ \bibinfo  {school} {Acta Universitatis Upsaliensis} (\bibinfo {year} {2012})\BibitemShut {NoStop}%
\bibitem [{\citenamefont {Wildes}\ \emph {et~al.}(2006)\citenamefont {Wildes}, \citenamefont {R{\o}nnow}, \citenamefont {Roessli}, \citenamefont {Harris},\ and\ \citenamefont {Godfrey}}]{wildes2006static}%
  \BibitemOpen
  \bibfield  {author} {\bibinfo {author} {\bibfnamefont {A.~R.}\ \bibnamefont {Wildes}}, \bibinfo {author} {\bibfnamefont {H.~M.}\ \bibnamefont {R{\o}nnow}}, \bibinfo {author} {\bibfnamefont {B.}~\bibnamefont {Roessli}}, \bibinfo {author} {\bibfnamefont {M.~J.}\ \bibnamefont {Harris}}, \ and\ \bibinfo {author} {\bibfnamefont {K.~W.}\ \bibnamefont {Godfrey}},\ }\bibfield  {title} {\enquote {\bibinfo {title} {Static and dynamic critical properties of the quasi-two-dimensional antiferromagnet {MnPS}$_3$},}\ }\href@noop {} {\bibfield  {journal} {\bibinfo  {journal} {Physical Review B—Condensed Matter and Materials Physics}\ }\textbf {\bibinfo {volume} {74}},\ \bibinfo {pages} {094422} (\bibinfo {year} {2006})}\BibitemShut {NoStop}%
\bibitem [{\citenamefont {Liu}\ and\ \citenamefont {Petrovic}(2018)}]{liu2018three}%
  \BibitemOpen
  \bibfield  {author} {\bibinfo {author} {\bibfnamefont {Y.}~\bibnamefont {Liu}}\ and\ \bibinfo {author} {\bibfnamefont {C.}~\bibnamefont {Petrovic}},\ }\bibfield  {title} {\enquote {\bibinfo {title} {Three-dimensional magnetic critical behavior in {CrI}$_3$},}\ }\href@noop {} {\bibfield  {journal} {\bibinfo  {journal} {Physical Review B}\ }\textbf {\bibinfo {volume} {97}},\ \bibinfo {pages} {014420} (\bibinfo {year} {2018})}\BibitemShut {NoStop}%
\bibitem [{\citenamefont {Dyson}(1956)}]{dyson1956thermodynamic}%
  \BibitemOpen
  \bibfield  {author} {\bibinfo {author} {\bibfnamefont {F.~J.}\ \bibnamefont {Dyson}},\ }\bibfield  {title} {\enquote {\bibinfo {title} {Thermodynamic behavior of an ideal ferromagnet},}\ }\href@noop {} {\bibfield  {journal} {\bibinfo  {journal} {Physical Review}\ }\textbf {\bibinfo {volume} {102}},\ \bibinfo {pages} {1230} (\bibinfo {year} {1956})}\BibitemShut {NoStop}%
\bibitem [{\citenamefont {Liu}\ and\ \citenamefont {Finkel'Stein}(2022)}]{liu2022spin}%
  \BibitemOpen
  \bibfield  {author} {\bibinfo {author} {\bibfnamefont {A.}~\bibnamefont {Liu}}\ and\ \bibinfo {author} {\bibfnamefont {A.~M.}\ \bibnamefont {Finkel'Stein}},\ }\bibfield  {title} {\enquote {\bibinfo {title} {Spin waves in layered antiferromagnets with honeycomb structure},}\ }\href@noop {} {\bibfield  {journal} {\bibinfo  {journal} {Physical Review B}\ }\textbf {\bibinfo {volume} {105}},\ \bibinfo {pages} {214409} (\bibinfo {year} {2022})}\BibitemShut {NoStop}%
\bibitem [{\citenamefont {Choe}\ \emph {et~al.}(2024)\citenamefont {Choe}, \citenamefont {Baydin}, \citenamefont {Ye}, \citenamefont {Ma}, \citenamefont {Nnokwe}, \citenamefont {Rodriguez-Vega}, \citenamefont {Chaudhary}, \citenamefont {Tay}, \citenamefont {He} \emph {et~al.}}]{choe2024magnetoelastic}%
  \BibitemOpen
  \bibfield  {author} {\bibinfo {author} {\bibfnamefont {J.}~\bibnamefont {Choe}}, \bibinfo {author} {\bibfnamefont {A.}~\bibnamefont {Baydin}}, \bibinfo {author} {\bibfnamefont {G.}~\bibnamefont {Ye}}, \bibinfo {author} {\bibfnamefont {B.}~\bibnamefont {Ma}}, \bibinfo {author} {\bibfnamefont {C.}~\bibnamefont {Nnokwe}}, \bibinfo {author} {\bibfnamefont {D.}~\bibnamefont {Rodriguez-Vega}, \bibfnamefont {M.and~Lujan}}, \bibinfo {author} {\bibfnamefont {S.}~\bibnamefont {Chaudhary}}, \bibinfo {author} {\bibfnamefont {F.}~\bibnamefont {Tay}}, \bibinfo {author} {\bibfnamefont {J.}~\bibnamefont {He}},  \emph {et~al.},\ }\bibfield  {title} {\enquote {\bibinfo {title} {Magnetoelastic coupling driven magnon gap in a honeycomb antiferromagnet},}\ }\href@noop {} {\bibfield  {journal} {\bibinfo  {journal} {Physical Review B}\ }\textbf {\bibinfo {volume} {110}},\ \bibinfo {pages} {104419} (\bibinfo {year} {2024})}\BibitemShut {NoStop}%
\bibitem [{\citenamefont {Huang}\ \emph {et~al.}(2022)\citenamefont {Huang}, \citenamefont {Kariyado},\ and\ \citenamefont {Hu}}]{huang2022topological}%
  \BibitemOpen
  \bibfield  {author} {\bibinfo {author} {\bibfnamefont {H.}~\bibnamefont {Huang}}, \bibinfo {author} {\bibfnamefont {T.}~\bibnamefont {Kariyado}}, \ and\ \bibinfo {author} {\bibfnamefont {X.}~\bibnamefont {Hu}},\ }\bibfield  {title} {\enquote {\bibinfo {title} {Topological magnon modes on honeycomb lattice with coupling textures},}\ }\href@noop {} {\bibfield  {journal} {\bibinfo  {journal} {Scientific Reports}\ }\textbf {\bibinfo {volume} {12}},\ \bibinfo {pages} {6257} (\bibinfo {year} {2022})}\BibitemShut {NoStop}%
\bibitem [{\citenamefont {Brehm}\ \emph {et~al.}(2024)\citenamefont {Brehm}, \citenamefont {Stagraczy{\'n}ski}, \citenamefont {Barna{\'s}}, \citenamefont {Dyrda{\l}},\ and\ \citenamefont {Qaiumzadeh}}]{brehm2024magnon}%
  \BibitemOpen
  \bibfield  {author} {\bibinfo {author} {\bibfnamefont {V.}~\bibnamefont {Brehm}}, \bibinfo {author} {\bibfnamefont {S.}~\bibnamefont {Stagraczy{\'n}ski}}, \bibinfo {author} {\bibfnamefont {J.}~\bibnamefont {Barna{\'s}}}, \bibinfo {author} {\bibfnamefont {A.}~\bibnamefont {Dyrda{\l}}}, \ and\ \bibinfo {author} {\bibfnamefont {A.}~\bibnamefont {Qaiumzadeh}},\ }\bibfield  {title} {\enquote {\bibinfo {title} {Magnon dispersion and spin transport in {C}r{C}l$_3$ bilayers under different strain-induced magnetic states},}\ }\href@noop {} {\bibfield  {journal} {\bibinfo  {journal} {Physical Review Materials}\ }\textbf {\bibinfo {volume} {8}},\ \bibinfo {pages} {054002} (\bibinfo {year} {2024})}\BibitemShut {NoStop}%
\bibitem [{\citenamefont {Chisnell}\ \emph {et~al.}(2015)\citenamefont {Chisnell}, \citenamefont {Helton}, \citenamefont {Freedman}, \citenamefont {Singh}, \citenamefont {Bewley}, \citenamefont {Nocera},\ and\ \citenamefont {Lee}}]{chisnell2015topological}%
  \BibitemOpen
  \bibfield  {author} {\bibinfo {author} {\bibfnamefont {R.}~\bibnamefont {Chisnell}}, \bibinfo {author} {\bibfnamefont {J.~S.}\ \bibnamefont {Helton}}, \bibinfo {author} {\bibfnamefont {D.~E.}\ \bibnamefont {Freedman}}, \bibinfo {author} {\bibfnamefont {D.~K.}\ \bibnamefont {Singh}}, \bibinfo {author} {\bibfnamefont {R.~I.}\ \bibnamefont {Bewley}}, \bibinfo {author} {\bibfnamefont {D.~G.}\ \bibnamefont {Nocera}}, \ and\ \bibinfo {author} {\bibfnamefont {Y.~S.}\ \bibnamefont {Lee}},\ }\bibfield  {title} {\enquote {\bibinfo {title} {Topological magnon bands in a kagome lattice ferromagnet},}\ }\href@noop {} {\bibfield  {journal} {\bibinfo  {journal} {Physical review letters}\ }\textbf {\bibinfo {volume} {115}},\ \bibinfo {pages} {147201} (\bibinfo {year} {2015})}\BibitemShut {NoStop}%
\bibitem [{\citenamefont {Riberolles}\ \emph {et~al.}(2024)\citenamefont {Riberolles}, \citenamefont {Slade}, \citenamefont {Han}, \citenamefont {Abernathy}, \citenamefont {Canfield}, \citenamefont {Ueland}, \citenamefont {Orth}, \citenamefont {Ke},\ and\ \citenamefont {McQueeney}}]{riberolles2024chiral}%
  \BibitemOpen
  \bibfield  {author} {\bibinfo {author} {\bibfnamefont {S.~X.~M.}\ \bibnamefont {Riberolles}}, \bibinfo {author} {\bibfnamefont {T.~J.}\ \bibnamefont {Slade}}, \bibinfo {author} {\bibfnamefont {B.}~\bibnamefont {Han}, \bibfnamefont {T.and~Li}}, \bibinfo {author} {\bibfnamefont {D.~L.}\ \bibnamefont {Abernathy}}, \bibinfo {author} {\bibfnamefont {P.~C.}\ \bibnamefont {Canfield}}, \bibinfo {author} {\bibfnamefont {B.~G.}\ \bibnamefont {Ueland}}, \bibinfo {author} {\bibfnamefont {P.~P.}\ \bibnamefont {Orth}}, \bibinfo {author} {\bibfnamefont {L.}~\bibnamefont {Ke}}, \ and\ \bibinfo {author} {\bibfnamefont {R.~J.}\ \bibnamefont {McQueeney}},\ }\bibfield  {title} {\enquote {\bibinfo {title} {Chiral and flat-band magnetic quasiparticles in ferromagnetic and metallic kagome layers},}\ }\href@noop {} {\bibfield  {journal} {\bibinfo  {journal} {Nature Communications}\ }\textbf {\bibinfo {volume} {15}},\ \bibinfo {pages} {1592} (\bibinfo {year} {2024})}\BibitemShut {NoStop}%
\bibitem [{\citenamefont {Seifert}\ and\ \citenamefont {Savary}(2022)}]{seifert2022phase}%
  \BibitemOpen
  \bibfield  {author} {\bibinfo {author} {\bibfnamefont {U.~F.~P.}\ \bibnamefont {Seifert}}\ and\ \bibinfo {author} {\bibfnamefont {L.}~\bibnamefont {Savary}},\ }\bibfield  {title} {\enquote {\bibinfo {title} {Phase diagrams and excitations of anisotropic {S}= 1 quantum magnets on the triangular lattice},}\ }\href@noop {} {\bibfield  {journal} {\bibinfo  {journal} {Physical Review B}\ }\textbf {\bibinfo {volume} {106}},\ \bibinfo {pages} {195147} (\bibinfo {year} {2022})}\BibitemShut {NoStop}%
\bibitem [{\citenamefont {Regnault}\ and\ \citenamefont {Rossat-Mignod}(1990)}]{regnault1990phase}%
  \BibitemOpen
  \bibfield  {author} {\bibinfo {author} {\bibfnamefont {L.~P.}\ \bibnamefont {Regnault}}\ and\ \bibinfo {author} {\bibfnamefont {J.}~\bibnamefont {Rossat-Mignod}},\ }\bibfield  {title} {\enquote {\bibinfo {title} {Phase transitions in quasi two-dimensional planar magnets},}\ }in\ \href@noop {} {\emph {\bibinfo {booktitle} {Magnetic Properties of Layered Transition Metal Compounds}}}\ (\bibinfo  {publisher} {Springer},\ \bibinfo {year} {1990})\ pp.\ \bibinfo {pages} {271--321}\BibitemShut {NoStop}%
\bibitem [{\citenamefont {Smirnova}\ \emph {et~al.}(2009)\citenamefont {Smirnova}, \citenamefont {Azuma}, \citenamefont {Kumada}, \citenamefont {Kusano}, \citenamefont {Matsuda}, \citenamefont {Shimakawa}, \citenamefont {Takei}, \citenamefont {Yonesaki},\ and\ \citenamefont {Kinomura}}]{smirnova2009synthesis}%
  \BibitemOpen
  \bibfield  {author} {\bibinfo {author} {\bibfnamefont {O.}~\bibnamefont {Smirnova}}, \bibinfo {author} {\bibfnamefont {M.}~\bibnamefont {Azuma}}, \bibinfo {author} {\bibfnamefont {N.}~\bibnamefont {Kumada}}, \bibinfo {author} {\bibfnamefont {Y.}~\bibnamefont {Kusano}}, \bibinfo {author} {\bibfnamefont {M.}~\bibnamefont {Matsuda}}, \bibinfo {author} {\bibfnamefont {Y.}~\bibnamefont {Shimakawa}}, \bibinfo {author} {\bibfnamefont {T.}~\bibnamefont {Takei}}, \bibinfo {author} {\bibfnamefont {Y.}~\bibnamefont {Yonesaki}}, \ and\ \bibinfo {author} {\bibfnamefont {N.}~\bibnamefont {Kinomura}},\ }\bibfield  {title} {\enquote {\bibinfo {title} {Synthesis, crystal structure, and magnetic properties of {B}i$_3${M}n$_4${O}$_12$({N}{O}$_3$) oxynitrate comprising s= 3/2 honeycomb lattice},}\ }\href@noop {} {\bibfield  {journal} {\bibinfo  {journal} {Journal of the American Chemical Society}\ }\textbf {\bibinfo {volume} {131}},\ \bibinfo {pages} {8313--8317} (\bibinfo {year} {2009})}\BibitemShut {NoStop}%
\bibitem [{\citenamefont {Seibel}\ \emph {et~al.}(2013)\citenamefont {Seibel}, \citenamefont {Roudebush}, \citenamefont {Wu}, \citenamefont {Huang}, \citenamefont {Ali}, \citenamefont {Ji},\ and\ \citenamefont {Cava}}]{seibel2013structure}%
  \BibitemOpen
  \bibfield  {author} {\bibinfo {author} {\bibfnamefont {E.~M.}\ \bibnamefont {Seibel}}, \bibinfo {author} {\bibfnamefont {J.~H.}\ \bibnamefont {Roudebush}}, \bibinfo {author} {\bibfnamefont {H.}~\bibnamefont {Wu}}, \bibinfo {author} {\bibfnamefont {Q.}~\bibnamefont {Huang}}, \bibinfo {author} {\bibfnamefont {M.~N.}\ \bibnamefont {Ali}}, \bibinfo {author} {\bibfnamefont {H.}~\bibnamefont {Ji}}, \ and\ \bibinfo {author} {\bibfnamefont {R.~J.}\ \bibnamefont {Cava}},\ }\bibfield  {title} {\enquote {\bibinfo {title} {Structure and magnetic properties of the $\alpha$-{N}a{F}e{O}$_2$-type honeycomb compound {N}a$_3${N}i$_2${B}i{O}$_6$},}\ }\href@noop {} {\bibfield  {journal} {\bibinfo  {journal} {Inorganic chemistry}\ }\textbf {\bibinfo {volume} {52}},\ \bibinfo {pages} {13605--13611} (\bibinfo {year} {2013})}\BibitemShut {NoStop}%
\bibitem [{\citenamefont {Klyushina}\ \emph {et~al.}(2017)\citenamefont {Klyushina}, \citenamefont {Lake}, \citenamefont {Islam}, \citenamefont {Park}, \citenamefont {Schneidewind}, \citenamefont {Guidi}, \citenamefont {Goremychkin}, \citenamefont {Klemke},\ and\ \citenamefont {M{\aa}nsson}}]{klyushina2017investigation}%
  \BibitemOpen
  \bibfield  {author} {\bibinfo {author} {\bibfnamefont {E.~S.}\ \bibnamefont {Klyushina}}, \bibinfo {author} {\bibfnamefont {B}~\bibnamefont {Lake}}, \bibinfo {author} {\bibfnamefont {A.~T. M.~N.}\ \bibnamefont {Islam}}, \bibinfo {author} {\bibfnamefont {J.~T.}\ \bibnamefont {Park}}, \bibinfo {author} {\bibfnamefont {A.}~\bibnamefont {Schneidewind}}, \bibinfo {author} {\bibfnamefont {T.}~\bibnamefont {Guidi}}, \bibinfo {author} {\bibfnamefont {E.~A.}\ \bibnamefont {Goremychkin}}, \bibinfo {author} {\bibfnamefont {B.}~\bibnamefont {Klemke}}, \ and\ \bibinfo {author} {\bibfnamefont {M.}~\bibnamefont {M{\aa}nsson}},\ }\bibfield  {title} {\enquote {\bibinfo {title} {Investigation of the spin-1 honeycomb antiferromagnet {B}a{N}i$_2${V}$_2${O}$_8$ with easy-plane anisotropy},}\ }\href@noop {} {\bibfield  {journal} {\bibinfo  {journal} {Physical Review B}\ }\textbf {\bibinfo {volume} {96}},\ \bibinfo {pages} {214428} (\bibinfo {year} {2017})}\BibitemShut {NoStop}%
\bibitem [{\citenamefont {Baithi}\ \emph {et~al.}(2023)\citenamefont {Baithi}, \citenamefont {Tran}, \citenamefont {Fix}, \citenamefont {Luong}, \citenamefont {Dhakal}, \citenamefont {Yoon}, \citenamefont {Rutkauskas}, \citenamefont {Kichanov}, \citenamefont {Zel} \emph {et~al.}}]{baithi2023incommensurate}%
  \BibitemOpen
  \bibfield  {author} {\bibinfo {author} {\bibfnamefont {N.~T.}\ \bibnamefont {Baithi}, \bibfnamefont {M.and~Dang}}, \bibinfo {author} {\bibfnamefont {T.~A.}\ \bibnamefont {Tran}}, \bibinfo {author} {\bibfnamefont {J.~P.}\ \bibnamefont {Fix}}, \bibinfo {author} {\bibfnamefont {D.~H.}\ \bibnamefont {Luong}}, \bibinfo {author} {\bibfnamefont {K.~P.}\ \bibnamefont {Dhakal}}, \bibinfo {author} {\bibfnamefont {D.}~\bibnamefont {Yoon}}, \bibinfo {author} {\bibfnamefont {A.~V.}\ \bibnamefont {Rutkauskas}}, \bibinfo {author} {\bibfnamefont {S.~E.}\ \bibnamefont {Kichanov}}, \bibinfo {author} {\bibfnamefont {I.~Y.}\ \bibnamefont {Zel}},  \emph {et~al.},\ }\bibfield  {title} {\enquote {\bibinfo {title} {Incommensurate antiferromagnetic order in weakly frustrated two-dimensional van der waals insulator {C}r{P}se$_3$},}\ }\href@noop {} {\bibfield  {journal} {\bibinfo  {journal} {Inorganic Chemistry}\ }\textbf {\bibinfo {volume} {62}},\ \bibinfo {pages} {12674--12682} (\bibinfo {year} {2023})}\BibitemShut {NoStop}%
\bibitem [{\citenamefont {Chern}()}]{giawei}%
  \BibitemOpen
  \bibfield  {author} {\bibinfo {author} {\bibfnamefont {G.~W.}\ \bibnamefont {Chern}},\ }\href@noop {} {\bibinfo  {journal} {Private communication}\ }\BibitemShut {NoStop}%
\bibitem [{\citenamefont {Kikuchi}\ \emph {et~al.}(2025)\citenamefont {Kikuchi}, \citenamefont {Ozeki}, \citenamefont {Kurita}, \citenamefont {Asai}, \citenamefont {Williams}, \citenamefont {Hong},\ and\ \citenamefont {Masuda}}]{kikuchi2025dirac}%
  \BibitemOpen
\bibfield  {journal} {  }\bibfield  {author} {\bibinfo {author} {\bibfnamefont {H.}~\bibnamefont {Kikuchi}}, \bibinfo {author} {\bibfnamefont {M.}~\bibnamefont {Ozeki}}, \bibinfo {author} {\bibfnamefont {N.}~\bibnamefont {Kurita}}, \bibinfo {author} {\bibfnamefont {S.}~\bibnamefont {Asai}}, \bibinfo {author} {\bibfnamefont {T.~J.}\ \bibnamefont {Williams}}, \bibinfo {author} {\bibfnamefont {T.}~\bibnamefont {Hong}}, \ and\ \bibinfo {author} {\bibfnamefont {T.}~\bibnamefont {Masuda}},\ }\bibfield  {title} {\enquote {\bibinfo {title} {Dirac magnon in honeycomb lattice magnet {N}i{T}i{O}$_3$},}\ }\href@noop {} {\bibfield  {journal} {\bibinfo  {journal} {Journal of the Physical Society of Japan}\ }\textbf {\bibinfo {volume} {94}},\ \bibinfo {pages} {024703} (\bibinfo {year} {2025})}\BibitemShut {NoStop}%
\end{thebibliography}%

\clearpage

\beginsupplement
\clearpage
\onecolumngrid  

\vspace*{2cm}  

\begin{center}
  {\Large \textbf{Supplementary Material: Magnetic dynamics in NiTiO$_3$ honeycomb antiferromagnet using neutron scattering}}\

\vspace{0.5 cm}

  {\normalsize Srimal Rathnayaka$^1$, Luke Daemen$^2$, Tao Hong$^2$, Songxue Chi$^2$, Stuart Calder$^2$, John A.~Schneeloch$^1$, Yongqiang Cheng$^2$, Bing Li$^2$, Despina Louca$^{1*}$}\

\vspace{0.3 cm}

  {\small $^1$Department of Physics, University of Virginia, Charlottesville, Virginia 22904, USA}\\
  {\small $^2$Neutron Scattering Division, Oak Ridge National Laboratory, Oak Ridge, Tennessee 37831, USA}
\end{center}

\vspace{1cm}  

\twocolumngrid

\section{Determination of the Magnetic Moments}

\begin{figure*}[b]
\begin{center}
	\includegraphics[width=0.8\textwidth]{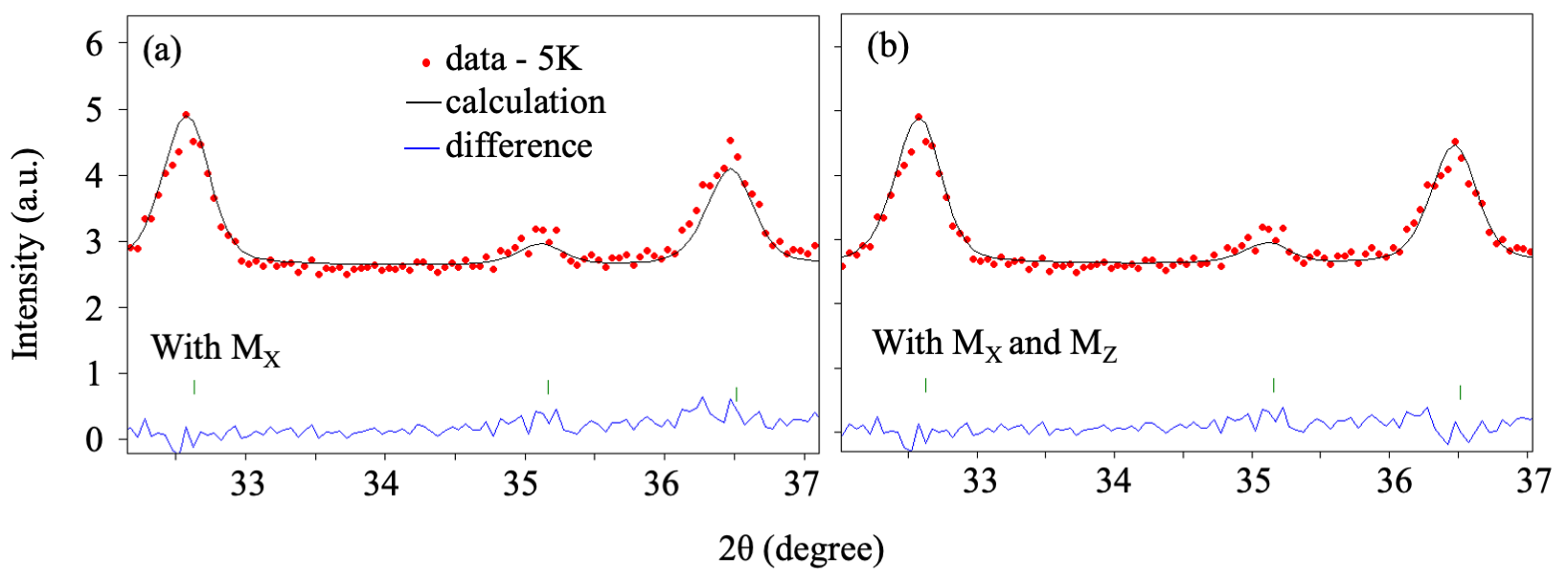}	
 \end{center}
	\caption{Rietveld refinement of the HB-2A data at 5 K using FullProf for the magnetic structure calculation: (a) refinement with magnetic moments constrained along the $x$ direction ($M_X$) only, and (b) refinement including both $x$ and $z$ components ($M_X$ and $M_Z$).} 
 \label{fig1}
\end{figure*}

In this section, refinement was performed on the HB-2A data at 5 K to investigate the presence of a possible out-of-plane magnetic moment in the system. Figure S1(a) shows the refinement using only an in-plane magnetic moment ($M_X$), while Fig. S1(b) presents the refinement including both in-plane ($M_X$) and out-of-plane ($M_Z$) components. The refinement incorporating both components yields a slightly better fit compared to the one using only $M_X$. In particular, the magnetic peak at 36.5$^\circ$ shows a noticeable improvement in fitting when both $M_X$ and $M_Z$ are included.

\section{Fitting of the linear spin wave model to the data.}

In this section, calculations were performed to assess the quality of the different J models with different J combinations. Fig. S2 shows the fitting of the Q-cut data using Gaussian fits, and Fig. S3 presents the extracted peak centers along with their corresponding experimental errors and the calculated dispersions from the linear spin wave theory. The goodness of fit (R-value) calculations were performed using fitspec method in the SpinW with the following equation:

\begin{equation}
   R = \sqrt{\frac{1}{n_E} \cdot \sum_{i,q} \frac{1}{\sigma_{i,q}^2} \left( E^{\text{sim}}_{i,q} - E^{\text{meas}}_{i,q} \right)^2 }
\end{equation}

Here, $E^{\text{meas}}_{i,q}$ represents the energy obtained from the peak fitting of the experimental data, $E^{\text{sim}}_{i,q}$ represents the energy obtained from the calculation, and ${\sigma_i^2}$ is the standard deviation obtained from the Gaussian fit of the experimental Q-cuts. ${i,q}$ indexing the spin wave mode and momentum respectively. 

\begin{figure*}[b]
\begin{center}
	\includegraphics[width=0.7\textwidth]{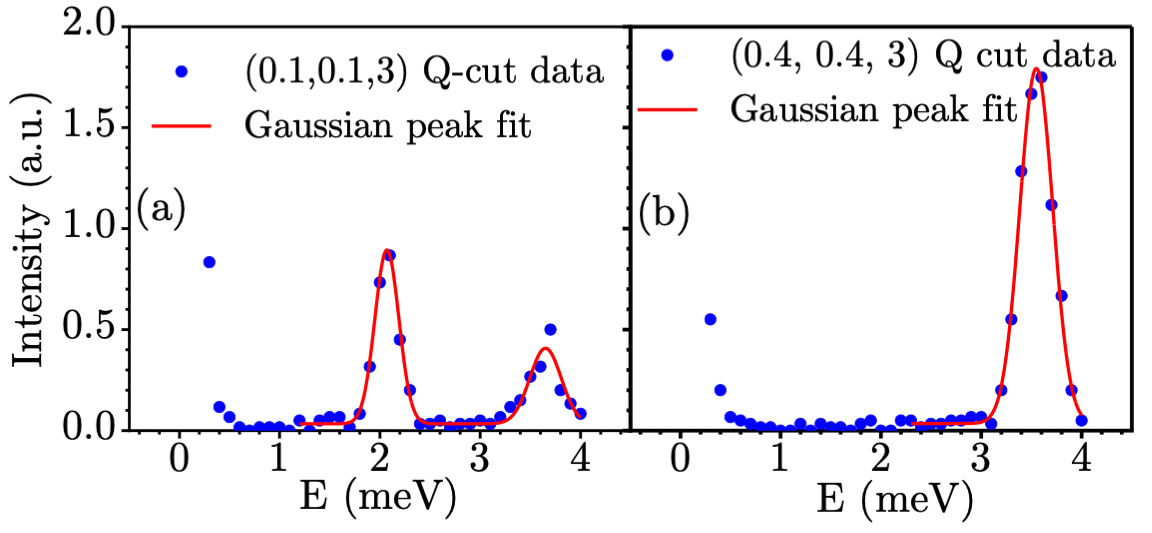}	
 \end{center}
	\caption{Gaussian fits for the Q-cuts of the (H,H,3) data along the (a) 0.1, 0.1, 3 and the (b) 0.4, 0.4, 3 directions. } 
 \label{fig2}
\end{figure*}

\begin{figure*}[b]
\begin{center}
	\includegraphics[width=1\textwidth]{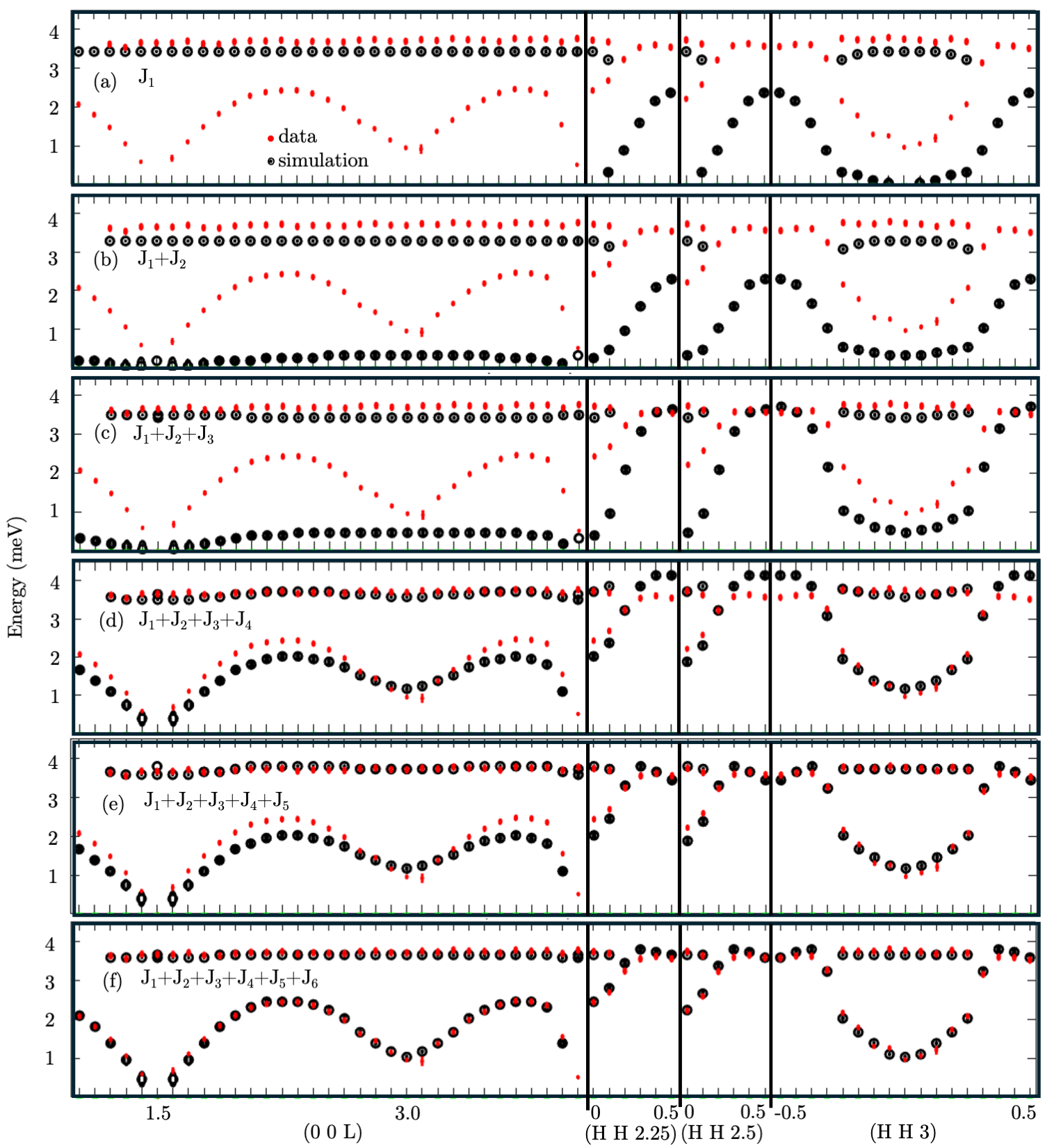}	
 \end{center}
	\caption{Comparison between the experimental data peaks and the error obtained from the Gaussian fits of the Q cuts as shown in Fig.S2 and the calculated best fitted linear spin wave models for different J combinations as shown in the figures. } 
 \label{fig3}
\end{figure*}

\begin{table}[t]

\caption{Exchange interactions for different the J models and corresponding R values.}
\centering
\setlength{\tabcolsep}{7pt} 
\begin{tabular}{c c c c c c c }
\hline\hline
         & 1J      & 2Js      & 3Js     & 4Js       & 5Js       & 6Js\\ [0.5 ex]
\hline
J1       & -0.55   & -0.50   & -0.50    & -0.28    & -0.17    & -0.12    \\
J2       & --      & 0.16    & 0.31     & 0.35     & 0.35     & 0.24     \\
J3       &  --     & --      & -0.15    & -0.13    & -0.10    & -0.08    \\
J4       &  --     & --      & --     & 0.25     & 0.25     & 0.25     \\
J5       &  --     & --      & --       & --    & -0.13    & -0.07    \\
J6       &  --     & --      & --       & --       & --       & 0.23     \\
D        & 0.20   &0.20    & 0.21    & 0.20    & 0.20    & 0.12   \\
R     & 131.4 & 121.8 & 116.8   & 35.6   & 33.0  & 13.3   \\ [1ex]
\hline
\end{tabular}
\label{table:nonlin}
\end{table}

In this analysis, the experimental data were fitted using linear spin wave theory, starting with a single J-value and progressively adding up to six J-values. Using the model and the data, R-values were calculated for each model. The corresponding R-values were calculated and are presented in Table I. The lower R-value, corresponding to the model with six J-values, confirms the accuracy and reliability of this model.

\section{Determination of the Curie–Weiss Temperature}

Magnetic susceptibility measurements were performed on the single-crystal NTO sample using a PPMS in the temperature range of 5–300 K. Fig.S4(a) shows the magnetization per Ni atom, calculated from the susceptibility data with temperature. The data exhibit AFM behavior in both FC warming and cooling conditions. Fig.S4(b) displays the inverse magnetization (1/M) from the FC cooling data. The data were fitted using the Curie–Weiss law, M = C/(T-$\theta$$_{CW}$), where C is the Curie constant.  The fit yields a $\theta_{CW}$ value of -43.48$\pm$0.32 K , indicating dominant antiferromagnetic interactions in the system.

\begin{figure*}[b]
\begin{center}
	\includegraphics[width=1\textwidth]{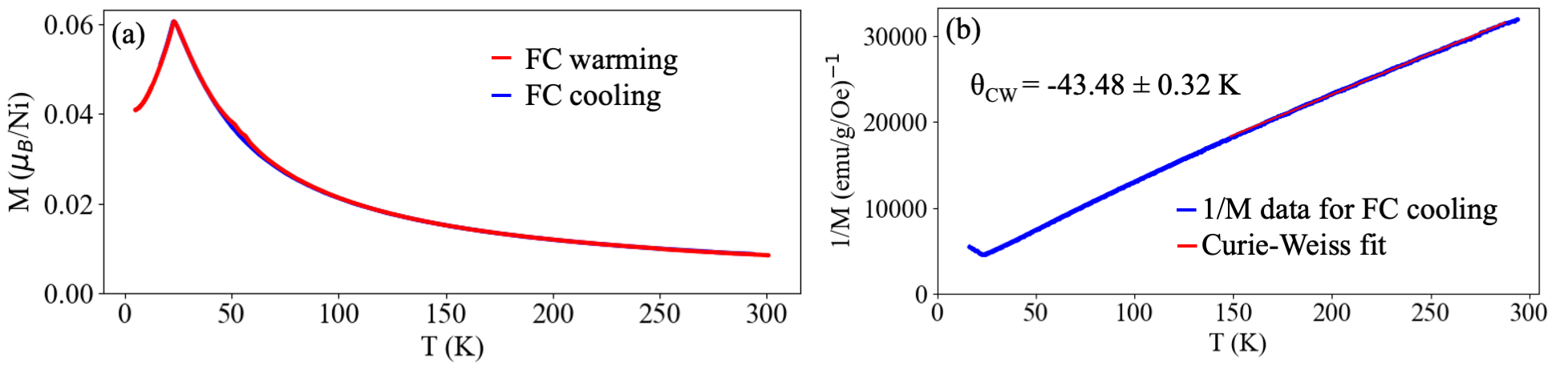}	
 \end{center}
	\caption{(a) FC magnetization of the single-crystal NTO sample measured during warming and cooling cycles. (b) Inverse magnetization (1/M) from the FC cooling data, fitted using the Curie–Weiss law to extract the effective magnetic moment and Curie–Weiss temperature.} 
 \label{fig4}
\end{figure*}

\end{document}